\documentclass[aps,pre,onecolumn,amsmath,superscriptaddress]{revtex4}
\usepackage{graphicx}
\usepackage{bm,color,textcomp}
\usepackage{amsmath,amsfonts,amssymb}

\usepackage{subfig}%per mettere le figure affiancate
\usepackage{graphicx}

\usepackage[normalem]{ulem}

\begin{document}
%%%%%%%%%%%%%%%%%%%%%%%%%%%%%%%%%%%%%%%%%%

\title{Active particles driven by competing spatially dependent self-propulsion and external force}

% Authors, for the paper (add full first names)
\author{Lorenzo Caprini} 
\affiliation{Heinrich-Heine-Universit\"at D\"usseldorf, Institut f\"ur Theoretische Physik II - Soft Matter, 
D-40225 D\"usseldorf, Germany.}
\author{Umberto Marini Bettolo Marconi}
\affiliation{School of Sciences and Technology, University of Camerino, Via Madonna delle Carceri, I-62032, Camerino, Italy.}
\author{Ren\'e Wittmann}
\affiliation{Heinrich-Heine-Universit\"at D\"usseldorf, Institut f\"ur Theoretische Physik II - Soft Matter, 
D-40225 D\"usseldorf, Germany.}
\author{Hartmut L\"owen}
\affiliation{Heinrich-Heine-Universit\"at D\"usseldorf, Institut f\"ur Theoretische Physik II - Soft Matter, 
D-40225 D\"usseldorf, Germany.}

\date{\today}

\begin{abstract}
We investigate how the competing presence of a nonuniform motility landscape and an external confining field affects the properties of active particles. 
%We employ the active Ornstein-Uhlenbeck particle (AOUP) model with a periodic swim velocity profile to derive analytical approximations for  the steady-state probability distribution of position and velocity. Our treatment encompasses both the Unified Colored Noise Approximation and the theory of potential-free active particles with spatially dependent swim velocity recently developed by us.
We employ the active Ornstein-Uhlenbeck particle (AOUP) model with a periodic swim velocity profile to derive analytical approximations for  the steady-state probability distribution of position and velocity, encompassing both the Unified Colored Noise Approximation and the theory of potential-free active particles with spatially dependent swim velocity recently developed.
%The theory is exact in the small-persistence limit, as shown by a multiple-time-scale expansion, which gives an expression of the phase-space distribution in terms of an infinite series in powers of the persistence time. In the large-persistence regime, our predictions still provide useful insight and we obtain qualitative agreement with the results of numerical simulations. 
%The theory is exact in the small-persistence limit, as shown by a multiple-time-scale expansion, while provides useful insight in the large-persistence regime.
%
%We test the theory by confining an active particle in a harmonic trap and modulating its self-propulsion velocity with a spatially periodic function. Even this simple setup, suggested by recent experiments, gives rise to interesting properties, which are not observed in the case of an AOUP with a constant swim velocity.
%In particular, we predict a transition from a unimodal to a bimodal spatial density, induced by decreasing the spatial period of the self propulsion (or increasing the persistence length),  which eventually leads to a multimodal shape. 
%Correspondingly, in the large-persistence regime, the velocity distribution shows pronounced deviations from the Gaussian shape, even displaying a bimodal profile in the high-motility regions.
%We further analyze this aspect, which emerges from the mutual correlation of position and velocity, by investigating numerically and analytically how the  local  kinetic temperature varies in space.
We test the theory by confining an active particle in a harmonic trap, which gives rise to interesting properties, such as a transition from a unimodal to a bimodal (and, eventually multimodal) spatial density, induced by decreasing the spatial period of the self propulsion. 
Correspondingly, the velocity distribution shows pronounced deviations from the Gaussian shape, even displaying a bimodal profile in the high-motility regions.
Our results can be confirmed by real-space experiments on active colloidal Janus particles in external fields.
\end{abstract}

\newcommand{\betadef}{\frac{1}{\tau}}
\newcommand{\alphadef}{\frac{\omega_q^2}{\gamma}}
\newcommand{\br}{{\bf r}}
\newcommand{\bu}{{\bf u}}
\newcommand{\bR}{{\bf x}}
\newcommand{\bRz}{{\bf x}^0}
\newcommand{\bk}{{ \bf k}}
\newcommand{\bx}{{ \bf x}}
\newcommand{\vv}{{\bf v}}
\newcommand{\nb}{{\bf n}}
\newcommand{\mb}{{\bf m}}
\newcommand{\bq}{{\bf q}}
\newcommand{\rb}{{\bar r}}

\newcommand{\eeta}{\boldsymbol{\eta}}
\newcommand{\xxi}{\boldsymbol{\xi}}

\maketitle

%%%%%%%%%%%%%%%%%%%%%%%%%%%%%%%%%%%%%%%%%%%%%%%%%%%%%%%%%%%%%%%%%%%%%%%%%
\section{Introduction}
%%%%%%%%%%%%%%%%%%%%%%%%%%%%%%%%%%%%%%%%%%%%%%%%%%%%%%%%%%%%%%%%%%%%%%%%%

The control of active matter~\cite{bechinger2016active, marchetti2013hydrodynamics, elgeti2015physics, gompper20202020} is an important issue for technological, 
biological and medical applications and has recently stimulated many experimental and theoretical works.
It is also very important in the future perspective of self-assembling and nano-fabricating active materials.
The motility of active particles is much higher than that of their passive counterparts and may be induced by either an internal ``motor'' (metabolic processes, chemical reactions, etc.)\ or an external driving force acting on each particle. 
%{\rene\it[how is the previous sentence related to the control aspect? Do we want to say that it is more difficult to control active particles than passive ones?? Then we should start the sentence by something like ``A major challenge arises from the fact that...'']}
This property offers intriguing perspectives since it is possible to achieve navigation control of active particles~\cite{lavergne2019group, sprenger2020active}, for instance driving their trajectories by some feedback mechanism~\cite{fernandez2020feedback, franzl2020active}.  

In the case of active colloids, such as Janus particles activated by external stimuli, the motility can be tuned by modulating the intensity of light~\cite{walter2007light, buttinoni2012active, palacci2013living, dai2016programmable, li2016light, uspal2019theory}.
This property has been employed to trap them~\cite{bregulla2014stochastic, jahanshahi2020realization} and to obtain polarization patterns induced by motility gradients~\cite{soker2021activity, auschra2021density}.
 Experimentally, the existence of an approximately linear relation between light intensity and swim velocity~\cite{lozano2016phototaxis} allows to tune the motility and design spatial patterns with specific characteristics.
Recent applications range from micro-motors~\cite{maggi2015micromotors, vizsnyiczai2017light} and rectification devices~\cite{stenhammar2016light, koumakis2019dynamic} to motility-ratchets~\cite{lozano2019propagating}.
Two experimental groups~\cite{arlt2018painting, arlt2019dynamics, frangipane2018dynamic},
 have devised a novel technique to control the swimming speed of bacteria by using patterned light fields to enhance/reduce locally their motility by increasing/decreasing the light intensity. 
This leads to a consequent accumulation/depletion of particles in some regions, 
so that this procedure can be used to draw two dimensional images with the bacteria~\cite{arlt2018painting}.

The fundamental physical concept behind experiments on light-controlled bacteria has been investigated
%Their experimental method is based on an idea introduced 
many years ago in a theoretical context for noninteracting random walkers by Schnitzer~\cite{schnitzer1993theory}
  and later been extended to the interacting Run-and-Tumble model by Cates and Tailleur~\cite{tailleur2008statistical}: the lower the speed of active particles, the higher their local density.
 This theoretical result has been tested and confirmed in many numerical works and the existence of such a relation between particle velocity and density is now considered one of the most distinguishing features of active matter.
%
%Concerning the theory,
A subsequent theoretical modeling of these effects has been proposed  in Refs.~\cite{lozano2016phototaxis, sharma2017brownian, ghosh2015pseudochemotactic, liebchen2019optimal, fischer2020quorum}  by generalizing the active Brownian particle (ABP) model to include a space-dependent swim velocity.
This additional ingredient accounts for the well-known quorum sensing~\cite{bauerle2018self, jose2021phase, azimi2020bacterial}, chemotaxis and  pseudochemotaxis~\cite{vuijk, merlitz, lapidus, vuijk2021} and correctly predicts a scaling of the density profile with the inverse of the swim velocity.
%The ABP model easily accommodates particle interactions, which 
Including particle interactions in the ABP model may lead to the spontaneous formation of a membrane in two-step motility profiles~\cite{grauer2018spontaneous} or cluster formation in regions with small activity~\cite{magiera2015trapping}.
Moreover, a temporal dependence in the activity landscape~\cite{maggi2018currents, merlitz2018linear, lozano2019diffusing, geiseler2017taxis,zampetaki2019taming, zhu2018transport}  may, in some cases, produce directed motion opposite to the propagation of the density wave~\cite{koumakis2019dynamic, geiseler2017self}. %These systems lead to coherent propagation of particle spikes~\cite{lozano2019propagating}, useful to separate binary mixtures~\cite{merlitz2018linear}, and, in some cases, produce counterintuitive directed motion opposite to the propagation of the density wave~\cite{koumakis2019dynamic, geiseler2017self}.

 In spite of its versatility and wide applicability~\cite{solon2015pressure, flenner2016nonequilibrium, petrelli2018active, caprini2021spatial, negro2022inertial, chacon2022intrinsic},
%(it can easily include interactions) and very intuitive description of the motion in terms of a combination of stochastic reorientation of the propulsion and particle propagation processes, 
the ABP model is harder to use to make theoretical progress than its ``sister/brother''~\cite{caprini2022parental}, the active Ornstein-Uhlenbeck particle (AOUP) model~\cite{maggi2015multidimensional, szamel2014self, dabelow2019irreversibility, berthier2019glassy, martin2021statistical,caprini2020inertial,nguyen2021active} . 
%In the former, the dynamics of the active force is a diffusive process on a manifold of constant radius, while in the latter each component of this force evolves as an Ornstein-Uhlenbeck process~\cite{martin2021statistical}.
In the former, the modulus of the active force is fixed and its orientation diffuses, while in the latter each component of the propulsion force evolves according to an Ornstein-Uhlenbeck process. %~\cite{martin2021statistical}. 
Therefore, AOUPs can be used as an alternative to ABPs with simplified dynamics~\cite{fily2012athermal, farage2015effective} but are also suitable to describe the dynamics of a colloidal particle in an active bath~ \cite{maggi2017memory, maes2020fluctuating}.
Moreover, a suitable mapping between the parameters of the two models can be performed on the level of the autocorrelation function of the self-propulsion velocity \cite{wittmann2017effective,footnote}, such that their predictions agree fairly well
for small and intermediate persistence time of the active motion~\cite{caprini2022parental}.
The present authors recently modified the AOUP model to include a space-dependent swim velocity in Ref.~\cite{caprini2021dynamics} 
and obtained exact results for both the density profile and velocity distribution of a potential-free particle. 
%The present authors have also considered an AOUP with space-dependent swim velocity and studied it in the absence of external force field obtaining exact results concerning the density distribution and the velocity distribution of the model under fairly general assumptions.

Analytical results for active particles in competing external potential and motility fields are sparse. 
Therefore, in this work, we extend the theoretical treatment by including the presence of an external force field,
revealing more interesting properties than those obtained for constant swim velocity. 
For example, the density profile exhibits multiple peaks due to a competition between external forces and motility patterns, while
the velocity distribution at fixed position displays a transition from a unimodal to a bimodal shape.
The paper is structured as follows:
In Sec.~\ref{sec:model}, we present the model to describe an active particle in a space-dependent swim velocity landscape and subject to an external potential, while, in Sec.~\ref{sec:Theoretical predictions}, we develop our theoretical approach to describe the steady-state properties of the system. The theory is numerically tested in Sec.~\ref{sec:harmonic} in the case of a harmonic potential and sinusoidal swim velocity profile.
Finally, conclusions and discussions are reported In Sec.~\ref{sec:conclusion}.
The appendices contain derivations and information supporting the theoretical treatment.

%to ease the notation for the general results of our AOUP model in an arbitrary spatial dimension $d$ dimensional case, we follow the convention of Ref.~\cite{caprini2021dynamics} and do not imply the mapping to the ABP model in our Eq. .
%%%%%% for any spatial dimension as in Ref. []. This is equivalent to diffusivity Da=v_0^2\tau.
%To make contact to the ABP model, our parameters would have to be rescaled by the additional dimensional factor $d$,
%to account for the mapping Da=v_0^2\tau/d, such that ... \sqrt{2\tau/d} in the last term of Eq
%In our results merely ammount to u(x) --> u(x)/sqrt{d}, i.e., same physics

%%%%%

\section{Model}\label{sec:model}

\subsection{Active particles with spatial-dependent swim velocity}

%The active particle model with a spatial-dependent swim velocity has been recently introduced in Ref.~\cite{caprini2021dynamics}.  
%This is a generalization of the active Ornstein-Uhlenbeck particles (AOUP) 51–56 commonly used to simplify the dynamics of active Brownian particles 57,58 but also to describe the behavior of colloidal particles immersed in active baths, for instance formed by bacteria.
To describe an active particle with spatially-dependent swim velocity, we employ the stochastic model introduced in Ref.~\cite{caprini2021dynamics},
representing a generalization of the AOUP dynamics.
%
%
%It represents a generalization of the active Ornstein-Uhlenbeck particles (AOUP)~\cite{maggi2015multidimensional, szamel2014self, wittmann2017effective, dabelow2019irreversibility, berthier2019glassy, martin2021statistical,caprini2020inertial,nguyen2021active} used as an alternative to the active Brownian particle model (ABP) to simplify its dynamics~\cite{fily2012athermal, farage2015effective} and suitable to describe system of active particles or colloidal particles immersed in active baths~ \cite{maggi2017memory, maes2020fluctuating}.
%to account for a nonuniform velocity landscape.
%
 The position of the active particle, $\mathbf{x}$, evolves according to overdamped dynamics 
supplemented by a stochastic equation for the active driving, the so-called self-propulsion (or active) force, $\mathbf{f}_a$.
Such a force term is responsible for the persistence of the trajectory and its physical origin depends on the system under consideration: flagella for bacteria and chemical reactions for Janus particles, to mention just two examples.
In the standard AOUP model, the active force $\mathbf{f}_a$ has the following form:
\begin{equation}
\label{eq:activeforcedefinition}
\mathbf{f}_a=\gamma v_0\boldsymbol{\eta} \,,
\end{equation}
where $\boldsymbol{\eta}$ is a two-dimensional Ornstein-Uhlenbeck process, $\gamma$ is the friction coefficient and $v_0$ is the constant swim velocity induced by the active force. %, which in the standard AOUP model is constant.
The generalization to a spatial-dependent swim velocity is achieved by the transformation $v_0\to u(\mathbf{x}, t)$ in Eq.~\eqref{eq:activeforcedefinition}, which introduces a dependence on both position and time.
The shape of $u(\mathbf{x}, t)$ must satisfy some properties related to physical arguments:
\begin{itemize}
\item[i)] positivity: $u(\mathbf{x}, t)\geq0$, for every $\mathbf{x}$ and $t$,
\item[ii)] boundedness: $u(\mathbf{x}, t)$ needs to be a bounded function of its arguments because the swim velocity cannot be infinite. 
\end{itemize}

Assuming inertial effects to be negligible at the microscopic scale, typically realized at small Reynolds numbers, the overdamped dynamics of the active particle with spatially modulating swim velocity reads: 
\begin{subequations}
\label{eq:AOUP}
\begin{align}
\label{eq:AOUP_x}
&\gamma\dot{\mathbf{x}}=\mathbf{F} + \gamma\sqrt{2 D_\text{t}}\boldsymbol{w}  + \gamma u(\mathbf{x}, t) \boldsymbol{\eta}\,, \\
\label{eq:AOUP_eta}
&\tau\dot{\boldsymbol{\eta}} = - \boldsymbol{\eta} + \sqrt{2\tau}\boldsymbol{\chi}  \,,
\end{align}
\end{subequations}
where $\boldsymbol{\chi}$ and $\boldsymbol{w}$ are $\delta$-correlated noises with zero average and unit variance  and $\mathbf{F}$ is the force exerted on the particle, resulting from the gradient of a potential $U(\mathbf{x})$, i.e., $\mathbf{F}(\mathbf{x})=-\nabla U(\mathbf{x})$. 
In this paper, we consider only a single particle, %noninteracting particles
so that $U(\mathbf{x})$ is a one-body potential, but the
description can be straightforwardly extended to the case of many interacting  particles. 
The coefficient $D_\text{t}$ is the translational diffusion coefficient due to the solvent satisfying the Einstein's relation with $D_\text{t}=T_t/\gamma$ and the temperature, $T_t$, of the passive bath (for unit Boltzmann constant). 
The dynamics of $\boldsymbol{\eta}$ is characterized by the typical time, $\tau$, which represents the correlation time of the active force autocorrelation and is usually identified with the persistence time of the single-trajectory, i.e., the time that a potential-free active particle spends moving in the same direction with velocity $u(\mathbf{x})$.
Finally, we remark that by taking $u(\mathbf{x}, t)=v_0$, the standard AOUP model is recovered upon substituting $v_0^2=D_\text{a}/\tau$, where $D_\text{a}$ is the diffusion coefficient due to the active force \cite{footnote}. 
Since $D_\text{t} \ll D_\text{a}$ in most of the experimental active systems~\cite{bechinger2016active}, in what follows, we neglect the contribution of the thermal bath by setting $D_\text{t}=0$.

\subsubsection{Velocity description of AOUP}

To ease the theoretical treatment of the model and gain further analytic insight, we switch to the auxiliary dynamics employed earlier in the potential-free case~\cite{caprini2021dynamics}. 
Instead of describing the system in terms of position $\mathbf{x}$ and self-propulsion velocity $u(\mathbf{x})\boldsymbol{\eta}$, we take advantage of the relation (holding for $D_\text{t}=0$)
\begin{equation}
\label{eq:relation_change}
\gamma\dot{\mathbf{x}}=\mathbf{F} + \gamma u(\mathbf{x}, t) \boldsymbol{\eta}
\end{equation}
to perform the simple change of variables $(\mathbf{x}, u(\mathbf{x},t)\boldsymbol{\eta}) \to (\mathbf{x}, \dot{\mathbf{x}}=\mathbf{v})$. 
This trick allows us to directly study the position and the velocity of the active particle as in the case $u(\mathbf{x},t)=v_0$.
As in the potential-free case, to return to the original variables, we need to account for the space-dependent Jacobian of the transformation given by the relation~\eqref{eq:relation_change}, which reads:
\begin{equation}
|J|=u(\mathbf{x},t) \,.
\end{equation}
This implies that the probability distributions, $\tilde{p}(\mathbf{x},\boldsymbol{\eta}, t)$ and $p(\mathbf{x}, \mathbf{v}, t)$, in the two coordinate frames satisfy the following relation:
\begin{equation}
\tilde{p}(\mathbf{x} ,\boldsymbol{\eta}, t) = |J|\, p(\mathbf{x}, \mathbf{v}, t)\,.
\end{equation}
In what follows, we use these new variables to study a system subject to both a
spatial-dependent swim velocity, $u(\mathbf{x}, t)$, and an external potential $U(\mathbf{x})$.
%In this paper, we extend the method to the case of spatial-dependent swim velocity, $u(\mathbf{x}, t)$, subject to an external potential.
%{\color{red} To proceed, we only consider the diffusion due to the active force, as stated above, which is usually some order of magnitude larger than the standard thermal diffusion. RIPETIZIONE INUTILE}
 The generalization to include a thermal noise can be achieved by following Ref.~\cite{caprini2018active}.

The leading steps to derive the auxiliary dynamics are illustrated in Appendix~\ref{appendix_1}, while here we only report the resulting equations:
\begin{subequations}
\label{eq:dynamics_vel}
\begin{align}
\label{eq:dynamics_xv}
\dot{\mathbf{x}}=&\mathbf{v}\,,\\
\label{eq:dynamics_vv}
\gamma\tau\dot{\mathbf{v}}=& - \gamma\Gamma(\mathbf{x}) \cdot\mathbf{v} - \nabla  U + \gamma u(\mathbf{x}, t) \sqrt{2\tau} \boldsymbol{\chi}\\
&+\tau\frac{\left[ \gamma\mathbf{v} + \nabla U \right]}{ u(\mathbf{x}, t)} \left( \frac{\partial}{\partial t} + \mathbf{v} \cdot \nabla \right) u(\mathbf{x}, t)  \,.\nonumber
\end{align}
\end{subequations}
In Eq.~\eqref{eq:dynamics_vv}, the first line is identical to the expression describing the constant case $u(\mathbf{x}, t)=v_0$: the dynamics of an overdamped active particle is mapped onto that of an underdamped passive particle with a space-dependent friction matrix, $\gamma \boldsymbol{\Gamma}(\mathbf{x})$, which depends on the second derivatives of the potential and reads:
\begin{equation}
\boldsymbol{\Gamma}(\mathbf{x})=\boldsymbol{\text{I}} +\frac{\tau}{\gamma}\nabla \nabla U (\mathbf{x}) \,,
\label{eq:Gamma}
\end{equation}
where $\mathbf{\text{I}}$ is the identity matrix.
Such a term increases or decreases the effective particle friction according to the value of the curvature of $U(\mathbf{x})$,
which becomes more and more important as $\tau$ becomes large.
In addition, as already found in the potential-free case, the noise amplitude contains a spatial and temporal dependence through the multiplicative factor $u(\mathbf{x}, t)$.
The second line of Eq.~\eqref{eq:dynamics_vv} contains the new terms, absent for $u(\mathbf{x})=v_0$, accounting for both  the time- and space-dependence of $u(\mathbf{x}, t)$.

For a further discussion of the new terms arising from a modulating swim-velocity profile, we
restrict ourselves to the time-independent case, $u(\mathbf{x}, t)=u(\mathbf{x})$.
Then, we identify two contributions to the total force. 
The first one, $\propto\mathbf{v} \mathbf{v} \cdot \nabla u$, is proportional to the square of the velocity and appears also in the absence of an external potential.
Since it is even under the time-reversal transformation, it cannot be interpreted as an effective Stokes force.
%but contributes to the acceleration experienced by a particle in a nonuniform motility landscape $u(\mathbf{x})$.
%
The second force, $\propto (\nabla U) \mathbf{v}\cdot \nabla u$, couples the gradients of the potential and the swim velocity
%The force i) 
%Instead, the force term ii) arises directly from the interplay between external potential and spatial dependent swim velocity. 
and gives rise to an extra space-dependent contribution to the effective friction. % proportional to the potential gradient and to the gradient of  the logarithm of $u(\mathbf{x})$.
 This allows us to absorb this term into an effective friction matrix $\boldsymbol{\Lambda}(\mathbf{x})$, which reads:
\begin{equation}
\label{eq:general_lambda_def}
\boldsymbol{\Lambda}(\mathbf{x}) = \boldsymbol{\Gamma}(\mathbf{x}) + \frac{\tau}{\gamma} \mathbf{F}(\mathbf{x})\frac{\nabla u(\mathbf{x})}{u(\mathbf{x})} \,,
\end{equation}
where $\boldsymbol{\Gamma}(\mathbf{x})$ is given by the expression for constant $u(\mathbf{x})=v_0$ (see Eq.~\eqref{eq:Gamma}).
%which contains a spatial dependence as far as $\tau/\gamma$ is large.
%{\color{red} As in the case $u(\mathbf{x})=v_0$, the term $\boldsymbol{\Lambda}(\mathbf{x})$ can be interpreted as a space-dependent friction force 
%that depends on $u(\mathbf{x})$ through the second term on the right-hand-side of Eq.~\eqref{eq:general_lambda_def}. 
The new term in Eq.~\eqref{eq:general_lambda_def} linearly increases with increasing $\tau$ and provides a further spatial dependence to the friction matrix.
Its sign is determined by $\nabla u(\mathbf{x}) $ and $\mathbf{F}(\mathbf{x})$, such that it can increase (positive sign) or decrease (negative sign) the effective friction. 
% with respect to the uniform case.
%As a matter of fact, the periodic profile of $u(x)$ splits the space in regions where $\Lambda(x)<\Gamma(x)$ and, viceversa, $\boldsymbol{\Lambda}(\mathbf{x})<\Gamma(x)$, slowing down or increasing the friction with respect to the homogeneous case.

\section{Theoretical predictions}
\label{sec:Theoretical predictions}

%{\rene To make analytical progress, we consider in the remainder of this work a system without thermal noise, $\boldsymbol{w}=\boldsymbol{0}$ or equivalently $D_\text{t}=0$, and}
In this section, we continue to
restrict ourselves to a static swim-velocity profile $u(\mathbf{x})$ to make theoretical progress. %which does not explicitly depend on time, $u(\mathbf{x}, t)=u(\mathbf{x})$.
At variance with the potential-free case, $U(\mathbf{x})=0$, the exact steady-state probability distribution  of positions and velocities, $p(\mathbf{x},\mathbf{v})$, is unknown and one needs to resort to some approximations. % to obtain a theoretical advance.
To this end,  we assume that all components of the probability current vanish, as in the case of a homogeneous swim velocity, $u(\mathbf{x})=v_0$. %, to find an approximate expression for the %velocity probability distribution  
As shown in Appendix~\ref{appendix_2}, %Appendix~\ref{appendix}, 
this condition means that in the Fokker-Planck equation associated to Eq.~\eqref{eq:dynamics_vel} the effective drift and diffusive terms mutually balance.
To derive a closed expression for the spatial density $\rho(\mathbf{x})$ we follow the idea of H\"anggi and Jung behind the Unified Colored Noise Approximation (UCNA)~\cite{jung1987dynamical,jung1988multi, hanggi1995colored}: 
one neglects the inertial term in the dynamics~\eqref{eq:dynamics_vv},
gets an effective overdamped equation for the particle position $\mathbf{x}$ and finally, via the associated Smoluchowski equation for $\rho(\mathbf{x},t)$,
obtains the stationary density distribution for a system with space-dependent activity.
The same $\rho(\mathbf{x})$ can be obtained using the path-integral method proposed by Fox~\cite{fox1986,fox1986b}.

 Here, we report only the main results while the details of the derivation can be found in Appendix~\ref{appendix_2}.
The whole stationary probability distribution reads:
\begin{equation}
\label{eq:powerexpansionintau}
\begin{aligned}
p(\mathbf{x},\mathbf{v}) \approx  \rho(\mathbf{x})\frac{\sqrt{\text{det}[\boldsymbol{\Lambda}(\mathbf{x})]}}{\sqrt{2\pi} u(\mathbf{x})} \exp{\left( -\frac{ \mathbf{v}\cdot\boldsymbol{\Lambda}(\mathbf{x})\cdot \mathbf{v}}{2 u^2(\mathbf{x})}  \right)}  \,,
\end{aligned}
\end{equation}
where $\text{det}[\cdot]$ represents the determinant of a matrix. 
We remark that the prefactor $\sqrt{\text{det}[\boldsymbol{\Lambda}(\mathbf{x})}]/(\sqrt{2\pi}u(\mathbf{x}))$ is the explicit factor normalizing the conditional velocity distribution (i.e., at fixed position $\mathbf{x}$).
The function $\rho(\mathbf{x})$ is approximated by 
\begin{equation}
\begin{aligned}
\label{eq:UCNA_generalized}
\rho(\mathbf{x}) \approx \frac{\mathcal{N}}{u(\mathbf{x})} \text{det}[\boldsymbol{\Lambda}(\mathbf{x})] \exp{\Biggl( \frac{1}{\gamma \tau} \int^x d\mathbf{y}\cdot \frac{\boldsymbol{\Lambda}(\mathbf{y}) \cdot\mathbf{F}(\mathbf{y}) }{u^2(\mathbf{y})}  \Biggr)} 
\end{aligned}
\end{equation}
with $\mathcal{N}$ being a normalization constant. Our expression for $\rho(\mathbf{x})$ coincides with the spatial density because it follows from integrating out the velocity in Eq.~\eqref{eq:powerexpansionintau}. 
The full distribution~\eqref{eq:powerexpansionintau} displays a multivariate Gaussian profile in the velocity, whose covariance
matrix accounts for the nontrivial coupling between velocity and position:
\begin{equation}
\label{eq:spatialvariance}
\langle \mathbf{v}\mathbf{v}(\mathbf{x}) \rangle = u^2(\mathbf{x})\boldsymbol{\Lambda}^{-1}(\mathbf{x}) \,.
\end{equation}
The covariance $\langle \mathbf{v}\mathbf{v}(\mathbf{x}) \rangle$ is spatially modulated by $u(\mathbf{x})$, which also occurs in the potential-free case, so that, in the regions where the swim velocity is large, the particle moves faster. 
In addition, the external potential not only affects the velocity covariance through $\Gamma(\mathbf{x})$, as in the case $u(\mathbf{x})=v_0$ (see for instance Refs.~\cite{marconi2016velocity, caprini2020active}), but contains an additional spatial dependence through the coupling to the velocity gradient in the second term of Eq.~\eqref{eq:general_lambda_def}. 
%term $\mathbf{F}(\mathbf{x})\nabla u(\mathbf{x})$. 
%Expression~\eqref{eq:powerexpansionintau} represents the best Gaussian approximation for the velocity distribution and generalizes the result obtained in the case $u(x)=v_0$ (Ref.~\cite{marconi2016velocity}) to a spatial-dependent motility landscape.}
%{\rene\it[Do we need this last sentence, we say this again below, and should we really claim it is the best??]}

Since the distribution %conditional velocity distribution is normalized, 
$\rho(\mathbf{x})$ from Eq.~\eqref{eq:UCNA_generalized} can be interpreted as the effective density distribution of the system,
the particle behaves as if it was subject to an effective potential, $\mathcal{V}(\mathbf{x}):=-\tau\gamma v_0^2\ln(\rho(\mathbf{x}))$, which explicitly reads:
\begin{equation}
\mathcal{V}(\mathbf{x})=-v_0^2 \int^x d\mathbf{y}\cdot \frac{\boldsymbol{\Lambda}(\mathbf{y}) \cdot\mathbf{F}(\mathbf{y}) }{u^2(\mathbf{y})} - \tau\gamma v_0^2 \ln{\left(v_0\frac{\text{det}[\boldsymbol{\Lambda}(\mathbf{y})]}{u(\mathbf{x})} \right)} \,,
\label{eq:Ueff}
\end{equation}
up to a constant.
This expression contains two terms, 
i) the spatial integral of the external force modulated by the inverse of the covariance matrix of the velocity distribution, cf.\ Eq.~\eqref{eq:spatialvariance}, %$u^2(\mathbf{x})\boldsymbol{\Lambda}^{-1}(\mathbf{x})$,
and ii) the logarithm containing both the velocity modulation $u(\mathbf{x})$ and the determinant of the spatially-dependent matrix $\boldsymbol{\Lambda}^{-1}(\mathbf{x})$. %, modulated by $u(\mathbf{x})$.
 Upon setting $u(\mathbf{x})=v_0$, we can perform the integral explicitly and $\mathcal{V}(\mathbf{x})$ reduces to the effective potential of  the UCNA, which is also equivalent to the result of the Fox approach since we neglect translational Brownian noise~\cite{wittmann2017effective}.
Note that the spatial dependence of the swim velocity gives rise to an additional potential term with respect to the case $u(\mathbf{x})=v_0$ contained in the expression for $\boldsymbol{\Lambda}(\mathbf{x})$.
%proportional to the scalar product of the gradient of the logarithm of $u(\mathbf{x})$ with $\mathbf{F}(\mathbf{x})$ (through the friction matrix $\boldsymbol{\Lambda}(\mathbf{x})$). 
At equilibrium, when $u(\mathbf{x})=v_0$ and $\tau \to 0$, the density reduces to the well known Maxwell-Boltzmann profile, since $\boldsymbol{\Lambda}(\mathbf{x})$ becomes constant.

\subsection{Multiscale method for the full-space distribution}

 To check the validity of our predictions, at least in the small-persistence regime, we resort to an exact perturbative approach in powers of the persistence time $\tau$.
For simplicity, the technique is presented in the one-dimensional case because the generalization to more dimensions is technically more involved and does not provide additional insight.
In addition, as in experiments based on active colloids \cite{lozano2016phototaxis}, we %anticipate that we 
will consider a one-dimensional swim velocity profile $u(x)$ in the remainder of this work,
justifying our particular attention to the one-dimensional case in the following presentation.

%{\rene\it[we used both names Fokker-Planck-Kramers and Fokker-Planck for the same equation. Which one should we keep?? I removed Kramers, you can change it all if you prefer! ]}

Our starting point is the following Fokker-Planck equation for the probability distribution $p(x,v,t)$:
\begin{flalign}
\label{eq:FokkerPlanck_potentialconfined}
\partial_t p = & \frac{\Lambda(x)}{\tau}\frac{\partial}{\partial v} \left( v p  \right) + \frac{u^2(x) }{\tau} \frac{\partial^2}{\partial v^2}   p - \frac{F(x)}{\tau\gamma} \frac{\partial}{\partial v} p      
-v \frac{\partial}{\partial_x} p -\frac{1}{u(x)}\left(\frac{\partial}{\partial x} u(x)\right) \frac{\partial}{\partial v} \left(v^2 p\right) \,,%\nonumber 
\end{flalign}
associated to the dynamics~\eqref{eq:dynamics_vel} in one spatial dimension.
Its solution
is unknown for a general potential $U(x)$, even in the special case $u(x)=v_0$.   
%At variance with the potential-free case, the steady-state probability distribution is unknown and 
Therefore, one needs to resort to approximation methods or perturbative strategies to obtain analytical insight. %expressions for $p(x,v)$.
As shown in previous work~\cite{bocquet1997high, marini2007theory},  it is possible to obtain 
perturbatively both the full distribution $p(x,v,t)$ and the configurational Smoluchowski equation
 for the reduced space distribution $\rho(x,t)$ following the method developed by Titulaer in the seventies~\cite{titulaer1978systematic}: 
 starting from the Fokker-Planck equation~\eqref{eq:FokkerPlanck_potentialconfined} % for $p(x,v,t)$
  the velocity degrees of freedom can be eliminated by using a multiple-time-scale technique.
  %to eliminate from the Fokker-Planck-Kramers equation for $p(x,v,t)$.
%The method allows a systematic derivation of the from the Kramers equation
%via the elimination of the velocity degrees of freedom.
% Such a procedure not only gives access to $\rho(x,t)$ but also allows to compute perturbatively the full $p(x,v,t)$ distribution. 
 %
 Physically speaking, the fast time scale of the system corresponds to the time interval necessary for the velocities of the particles to relax to the configurations consistent with the values imposed by the vanishing of the currents. 
The characteristic time of the slow time scale is much longer and corresponds to the time necessary for the positions of the particles to relax towards the stationary configuration.
  
 %The Fokker-Planck equation for the probability distribution, $p(v, x, t)$, associated to the dynamics~\eqref{eq:stochastic_potential} is straightforwardly derived and reads:
%\begin{flalign}
%\label{eq:FokkerPlanck_potentialconfined}
%\partial_t p = & \frac{\Lambda(x)}{\tau}\partial_v \left[ v p  \right] + \frac{u^2(x) }{\tau} \partial^2_{v^2}   p - \frac{F(x)}{\tau\gamma} \partial_v p\\
%&-v \partial_x p -\frac{1}{u(x)}(\partial_x u(x)) \partial_v \left[v^2 p\right] \,.\nonumber 
%\end{flalign}
%At variance with the potential-free case, the steady-state probability distribution is unknown and one needs to resort to approximation methods or perturbative strategies to obtain the form of $p(x,v)$.
 
 In the present case, the perturbative parameter is the persistence time $\tau$. 
Since we are mainly interested in time-independent properties, we limit ourselves to compute the steady-state probability distribution by generalizing the results of Ref.~\cite{fodor2016far,marconi2017heat} %by generalizing the method 
previously obtained for the case $u(x)=v_0$ (see also Ref.~\cite{martin2021aoup} for a more general expansion with an additional thermal noise).
 For space reasons, the details of the calculations are reported in Appendix~\ref{appendix}. %, while the main result are summarized below. %in this Section.
 Our main result is
%In particular, expanding the result in powers of the parameter $\tau$ we obtain 
the following exact perturbative expansion of the distribution $p(x,v)$ in powers of the parameter $\tau$:~\cite{footnote2}
\begin{equation}
\label{eq:pxv_pertubativetau}
\begin{aligned}
p(x,v) = &\rho_s(x) p_s(x,v) \Biggl\{ 1+\frac{\tau}{\gamma} \Bigl[ \frac{1}{2} U''(x) - \frac{v^2}{2u^2(x)} U''(x)  +\left( \frac{v^2}{2u(x)^2}-\frac{1}{2} \right) U'(x) \frac{u'(x) }{u(x)} \Bigr] \\
&+\frac{\tau^2}{6\gamma} u(x) \left( \frac{v^3}{u^3(x)}-3\frac{v}{u(x)}  \right)\left[ U'''(x)  -  \frac{\partial}{\partial x} \left(\frac{U'(x)}{u(x)}u'(x) \right) \right] \Biggr\}+O(\tau^3)
\end{aligned}
\end{equation}
where the normalized distribution $p_s(x,v)$ is given by %a Maxwellian distribution with space-dependent variance 
\begin{equation}
\label{eq:potentialfree_solution}
p_s(x,v) = \frac{\mathcal{N}}{\sqrt{2\pi} u(x)} \exp{\left( - \frac{v^2}{2u^2(x)}  \right)} \,,
\end{equation}
and the function $\rho_s(x)$ reads
%{\rene\it[maybe call it $\rho_s(x)$ if it is not equal to the exact $\rho(x)$!]}
\begin{equation}
\rho_s(x)=\mathcal{N}\frac{\Lambda(x)}{u(x)}\exp{\left( - \frac{1}{\gamma\tau}\int^x dy\, U'(y) \frac{\Lambda(y)}{u^2(y)}  \right)} \,.
\label{eq:diciassette}
\end{equation}
with the normalization factor $\mathcal{N}$  and the prime as a short notation for the spatial derivative.
Already at order $\tau/\gamma$ our general result~\eqref{eq:pxv_pertubativetau} for a nonuniform swim velocity contains an extra term proportional to $\partial_x u(x)$, compared to the expansion derived in Ref.~\cite{fodor2016far, marconi2017heat,footnote2}, which is responsible for an additional coupling between position and velocity.

The product $\rho_s(x) \times p_s(x,v)$ in Eq.~\eqref{eq:pxv_pertubativetau} plays the role of an effective equilibrium-like distribution, which is exact in the limit $\tau\to0$. 
The required expression~\eqref{eq:potentialfree_solution} for $p_s(x,v)$ 
is the exact solution of the potential-free active system with spatial-dependent swim velocity as derived in Ref.~\cite{caprini2021dynamics}: 
it is a Gaussian probability distribution for the particle velocity $v$ with an effective space-dependent kinetic temperature provided by $u^2(x)$.
The spatial density $\rho_s(x)$ from Eq.~\eqref{eq:diciassette} corresponds to the UCNA result~\eqref{eq:UCNA_generalized} in one dimension.
 Our previous approximated expression~\eqref{eq:powerexpansionintau} for $p(x,v)$ is consistent with
the full result~\eqref{eq:pxv_pertubativetau} 
%Indeed, the density distributions $\rho(x)$ in two cases, Eqs.~\eqref{eq:UCNA_generalized} and~\eqref{eq:diciassette}, coincide. 
%Regarding the two expressions~\eqref{eq:pxv_pertubativetau} and~\eqref{eq:powerexpansionintau} they coincide 
at first order in the expansion parameter $\propto \tau$, while
 the first deviation between the two formulas occurs at order $O(\tau^2)$, where the exact expression for $p(x,v)$ also contains odd terms in $v$.
Thus,  the exact density profile $\rho(x)=\int\mathrm{d}vp(x,v)=\rho_s(x)+O(\tau^2)$ deviates from the UCNA result beyond linear orders in $\tau$.

\section{The harmonic oscillator}\label{sec:harmonic}

%---------------------------- Fig.1 ----------------------------------
\begin{figure*}[t!]
\includegraphics[width=0.8\textwidth]{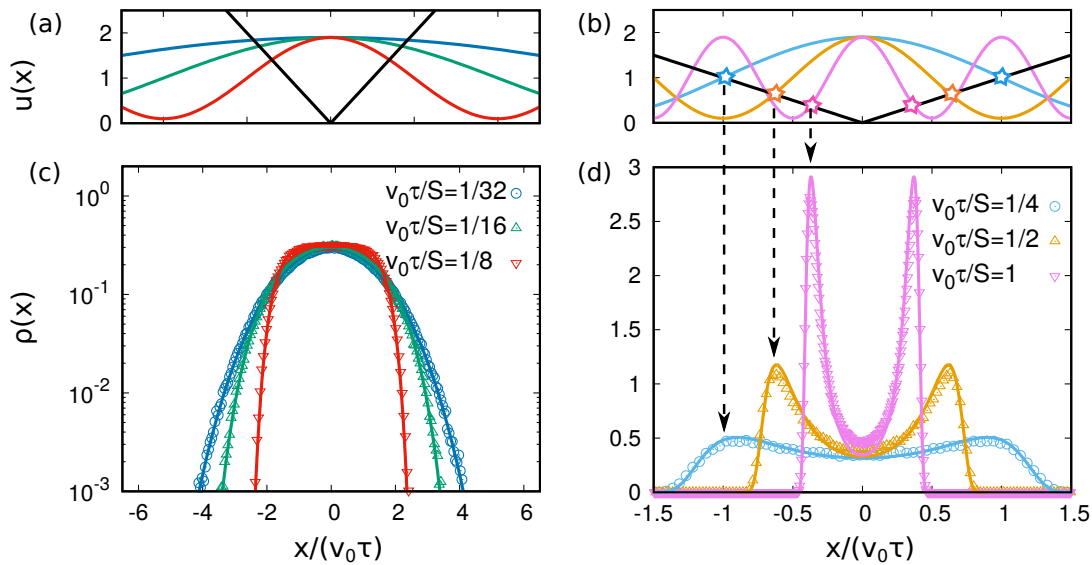}
\caption{
Density distributions. Panels (a) and (b): swim velocity profile $u(x)$ for different values of $S$ (colored curves), compared with the linear force profile, $F(x)/\gamma=-k x /\gamma$ (black curve). The colored stars are placed at the first cross point between $u(x)$ and $F(x)/\gamma$.
Panels (c) and (d): spatial density profile, $\rho(x)$, for different values of $S$. Points are obtained by numerical simulations while solid lines by plotting the theoretical prediction~\eqref{eq:UCNA_generalized}.
Panels (a), (c) and (b), (d) share the same legend. Simulations are realized with $\tau k/\gamma=1$ and $\alpha=0.9$.
 }
\label{fig:density_smallintermediateS}
\end{figure*}
%---------------------------------------------------------------------

In this section, we present and investigate the interplay between a spatially modulated swim velocity and an external confining potential in one spatial dimension.
While, in Sec.~\ref{sec:Theoretical predictions}, we have shown that our analytical predictions from Eqs.~\eqref{eq:powerexpansionintau},~\eqref{eq:UCNA_generalized} and~\eqref{eq:spatialvariance} are exact in the small-persistence regime through analytical arguments, a numerical analysis is necessary to check our approximations in the large-persistence regime.

To fix the form of the profile $u(x)$ employed in our numerical study and theoretical treatment,
we take inspiration from experimental works on active colloids~\cite{lozano2016phototaxis} and
 consider a static periodic profile $u(x)$ varying along a single direction, namely the $x$ axis, so that:
\begin{equation}
\label{eq:swimvelocityprofile}
u(x)=v_0\left(1+\alpha \cos{\left( 2\pi \frac{x}{S}\right)} \right)\,,
\end{equation}
where $\alpha<1$ and $v_0>0$ so that $u(x)>0$ for every $x$. 
The parameter $\alpha$ determines the amplitude of the swim velocity oscillation while $S>0$ sets its spatial period. 
As a consequence, the active particle is subject to the minimal swim velocity $v_0(1-\alpha)$ and to the maximal one $v_0(1+\alpha)$.
This choice allows us to look only at the $x$ component of the system, reducing the dimensionality of the problem. 
%and {\color{red} LA FRASE E' TRONCATA}\\

%{\rene Before discussing the effect of introducing the confining force, $F(x)=-U'(x)$,}
We remind that, in the potential-free case~\cite{caprini2021spatial}, the system admits two typical length scales, i.e., the persistence length $v_0 \tau$ and the spatial period, $S$, of the swim velocity profile~\eqref{eq:swimvelocityprofile}. 
In other words, by rescaling the time by $\tau$ and the particle position by $v_0\tau$, the dynamics is controlled by the dimensionless parameter $v_0 \tau/S$ and by the dimensionless parameter $\alpha$ quantifying the amplitude of the swim velocity oscillation. 
The external force $F(x)$ then introduces at least one additional length-scale, $\ell$, which depends on the specific nature of $F$, and, thus, an additional dimensionless parameter, say $\ell/(v_0\tau)$, related to the external potential.  
The last dimensionless parameter controls the dynamics also in the case $u(x)=v_0$~\cite{caprini2019activity}. 
Now, we can identify the small-persistence regime, where the self-propulsion velocity relaxes faster than the particle position, with the criterion $v_0 \tau/S \ll 1$ and $\ell/(v_0\tau) \gg 1$. 
Under the former condition, we expect that the system behaves as its passive counterpart: if $\tau \ll S/v_0$ holds, the self-propulsion behaves as an effective white noise with amplitude. 
In the opposite case, when $v_0 \tau/S \gg 1$, % and $\ell/(v_0\tau) \ll 1$, 
the dynamics is strongly persistent and we expect intriguing nonequilibrium properties.
%{\rene which include effects of the external potential if, in addition, $\ell/(v_0\tau) \ll 1$ holds}.
%{\rene\it[we never use $\ell$ in the following, but we should make contact to the above discussion when introducing the harmonic potential!]}

Now, we confine the particles through a harmonic potential, 
\begin{equation}
U(x)=\frac{k}{2} x^2 \,,
\end{equation}
where the constant $k$ determines the strength of the linear force. 
The dimensionless parameter associated with this external potential is thus $k \tau/ \gamma$, i.e., $\ell=v_0 \tau^2 k/\gamma$, and the effective friction coefficient $\Lambda(x)$ from Eq.~\eqref{eq:general_lambda_def} becomes:
\begin{equation}
\label{eq:lambda_harmonic}
%\Lambda(x)=1+\tau \frac{k}{\gamma} +\alpha \tau \frac{k}{\gamma} \frac{x}{u(x)^2} \frac{2 \pi}{S}\sin{\left(2\pi \frac{x}{S}\right)}
\Lambda(x)=\left(1+\tau \frac{k}{\gamma}\right) \left(  1+  \frac{ \alpha x}{u(x)^2} \frac{2 \pi}{S}\sin{\left(2\pi \frac{x}{S}\right)}\frac{\tau \frac{k}{\gamma}}{1+\tau \frac{k}{\gamma}}\right)
\end{equation}
since the curvature of the potential is constant.
As shown by Eq.~\eqref{eq:lambda_harmonic}, the two dimensionless parameters $\tau k/\gamma$ and $\alpha$ play a similar role. 
 Indeed, they only determine the relative amplitude of the spatial modulation of $\Lambda(x)$. 
When either $\alpha$ or $\tau k/\gamma$ vanish, the effective friction becomes constant and the coupling between velocity and position disappears. 
Instead, when both $\tau k/\gamma \to\infty$ and $\alpha$ are increased ($\alpha\to1$),
the amplitude spatial oscillations becomes maximal.
%the ratio between the spatially-dependent and the  constant  term approaches its maximum value.
By varying the dimensionless parameter $v_0\tau/S$, on the other hand, one can explore the different properties of the system: when $v_0\tau/S$ grows, the spatial period of $u(x)$ increases and
the spatially-dependent term of $\Lambda(x)$ becomes less relevant.
 To study the resulting behavior of the system in detail, we keep fixed $\alpha=0.9$ and $\tau k/\gamma=1$ %$\alpha$ and $\tau k/\gamma$ 
and we change only $v_0\tau/S$ to study the properties of the system.

\subsection{Density distribution}\label{subsec:density}

\begin{figure*}[t!]
\includegraphics[width=0.8\textwidth]{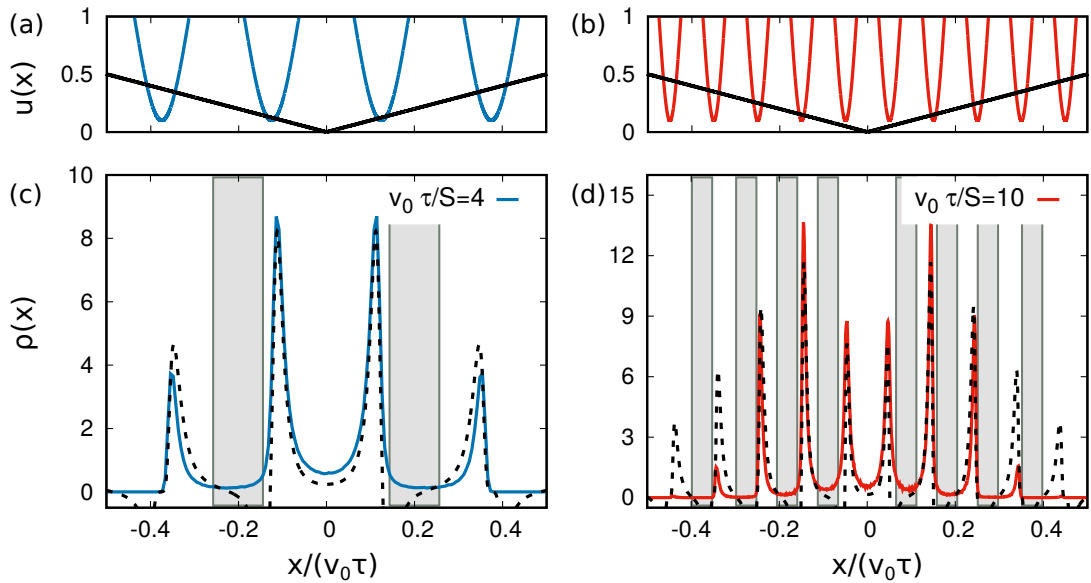}
\caption{
Density distributions. Panels (a) and (b): swim velocity profile $u(x)$ for two different values of $S$ (colored curves), compared with the linear force profile, $F(x)=-k x/\gamma$ (black curve). 
Panels (c) and (d): spatial density profile, $\rho(x)$, for different values of $S$. Solid colored lines (blue in panel (c) and red in panel (d)) are obtained by numerical simulations while dashed black lines by plotting the theoretical prediction~\eqref{eq:UCNA_generalized}.
Grey rectangles are drawn in the regions where Eq.~\eqref{eq:UCNA_generalized} is not defined, say when $\Lambda(x)<0$ (see the main text for more details).
Panels (a), (c) and (b), (d) share the same legend. Simulations are realized with $\tau k/\gamma=1$ and $\alpha=0.9$.
 }
\label{fig:densityCosinus_smallS}
\end{figure*}

We first focus on the spatial density profile, $\rho(x)$, shown in the bottom panels of Figs.~\ref{fig:density_smallintermediateS} and~\ref{fig:densityCosinus_smallS} 
for different  values of the spatial period of $u(x)$ (through the dimensionless parameter $v_0\tau/S$), reported in the top panels of Figs.~\ref{fig:density_smallintermediateS} and~\ref{fig:densityCosinus_smallS}.

%Figures~\ref{fig:density_smallintermediateS} (c) and (d) show the density profile, $\rho(x)$, for different values of $v_0\tau/S$ while Fig.~\ref{fig:density_smallintermediateS} (a) and (b) report the corresponding shapes of the swim velocity $u(x)$.
In the small-persistence regime, $v_0\tau/S \ll 1$ (see panels (a) and (c) of Fig.~\ref{fig:density_smallintermediateS}), the unimodal  density distribution is fairly described by % a, whose profile can be achieved by 
expanding the UCNA solution~\eqref{eq:UCNA_generalized} in powers of $x/S$, obtaining:
\begin{equation}
\label{eq:approx_rho_expansion}
\rho(x)\sim \exp{\left( -\frac{1+\tau \frac{k}{\gamma}}{(1+\alpha)^2} \frac{k}{v_0^2}\frac{ x^2}{2}  \right)} \,.
\end{equation}
In the expression~\eqref{eq:approx_rho_expansion}, we have neglected the terms proportional to $x^2/S^2$, $x^4/S^2$ and all higher-order terms in power of $\sim 1/S$.
Remarkably, even in this crude approximation, we see from the factor $(1+\alpha)^2$ that %case, 
the oscillations of the swim velocity 
lead to a decrease of the second moment $\langle x^2\rangle$ of $\rho(x)$ compared  
to the homogeneous case $u(x)=v_0$. 
This prediction is consistent with previous results obtained in the absence of an external potential, 
where the swim-velocity oscillations produce the decrease of the long-time diffusion coefficient~\cite{caprini2021dynamics} (see also Ref.~\cite{breoni2021brownian}).
In this regime, the spatial pattern $u(x)$ produces an effective potential with increasing stiffness for increasing spatial modulation. %stiffer than the one with $u(x)=v_0$.
For higher $v_0\tau/S$, the distribution starts developing non-Gaussian tails, which are still well-described by including higher-order terms in the UCNA expansion~\eqref{eq:approx_rho_expansion}. 

When increasing $v_0\tau/S$ further (see panels (b) and (d) of Fig.~\ref{fig:density_smallintermediateS}), $\rho(x)$ becomes a bimodal distribution with two peaks symmetric to the origin,
as in a system confined in a double-well potential. % (see panel (d)).
This effect is absent in the case $u(x)=v_0$ where the AOUP density distribution in a harmonic potential always has a Gaussian shape~\cite{szamel2014self, das2018confined, caprini2019comparative, dabelow2021irreversible}.
For a space-dependent swim velocity, the comparison between the analytical result~\eqref{eq:UCNA_generalized} and the numerical simulations still reveals a good agreement: % with data: 
in particular, Eq.~\eqref{eq:UCNA_generalized} is able to predict the observed bimodality of the distribution. % occurring at some critical values of the parameters. % (see the comparison between dashed and solid lines in panel (d)). 
To explain the occurrence of this bimodality in the shape of $\rho(x)$, we can use an effective (but rather general) %an effective 
force-balance argument in Eq.~\eqref{eq:AOUP_x}. 
This argument can be applied to %makes sense only in 
 the present intermediate-persistence regime, $v_0\tau/S\sim1$ (or also for $v_0\tau/S\gg1$ discussed later), 
where the self-propulsion vector $\eta$ in the active force can be considered to be roughly constant for typical times $t\lesssim \tau$.
Since the variance of $\eta$ is unitary, the most likely value assumed by the self-propulsion velocity at point $x$ is simply $u(x)$ (in absolute value). 
%{\rene \it[what exactly do you mean here? Isn't $u(x)$ rather the square root of the mean-squared self-propulsion velocity at point $x$?]}.
For this reason, it is generally unlikely to find the particle in regions with $u(x)<F(x)/\gamma$, because there the particle's self propulsion is not sufficient to climb up the potential gradient.
Moreover, in the spatial points where $u(x)>F(x)/\gamma$,
the active particle does not get %cannot remain
stuck on average because its high self-propulsion velocity allows its directed motion  until $u(x)=F(x)/\gamma$ is fulfilled. 
When this force balance occurs, the particle can explore further spatial regions only because of large (and rare) fluctuations of $\eta$.
%{\rene (which are only sufficient for the particle to enter regions with $u(x)>F(x)/\gamma$ for small periodicities $S\ll v_0\tau$)}. 
This reasoning is confirmed by inspecting Fig.~\ref{fig:density_smallintermediateS} for different fixed values of $v_0\tau/S$. %(c), reporting $u(x)$, and~\ref{fig:density_smallintermediateS}(d), displaying $\rho(x)$,  for the  same values of $S$ {\rene (dashed arrows)}. 
 It is evident from the dashed arrows that the peaks of the distribution in Fig.~\ref{fig:density_smallintermediateS}(d) coincide with the intersection between the external force $F(x)/\gamma$ (black curve) and $%v_0 
 u(x)$ (colored curves) in Fig.~\ref{fig:density_smallintermediateS}(c).

%---------------------------- Fig.2 ----------------------------------
%---------------------------------------------------------------------

Starting from the theoretical result~\eqref{eq:UCNA_generalized}, we can predict the critical value $S_c$ at which the distribution becomes bimodal, by simply requiring that $\mathrm{d}^2/\mathrm{d}x^2 \rho(x)=0$, obtaining:
\begin{equation}
\frac{S_c^2}{v_0^2 \tau^2}= (2\pi)^2 \alpha(1+\alpha)\left[\frac{1+3\tau \frac{k}{\gamma} }{\left(1+\tau \frac{k}{\gamma}\right)^2}\right] \frac{\gamma}{k\tau} \,.
\label{eq:Sc}
\end{equation}
In general, we predict that the value of $S_c/(v_0\tau)$ increases with $\alpha$ (we remind that $0<\alpha<1$) and is a decreasing function of $\tau k/\gamma$.
This is consistent with our physical intuition: larger oscillations  (i.e., larger $\alpha$) facilitate the transition to a bimodal shape. Indeed, the larger $\alpha$, the smaller the minimal self-propulsion velocity, that hinders the particle ability to come back to the origin.
Instead, the increase of $\tau k/\gamma$ gives rise to the opposite behavior: the larger $\tau k/\gamma$, the steeper the effective confining trap. As a consequence,  the  active particle needs larger fluctuations of $\eta$ to reach spatial regions where $u(x)$ assumes low values which compete with the external force.
Specifically, for the chosen parameters $\alpha=0.9$ and $\tau k/\gamma=1$, Eq.~\eqref{eq:Sc} predicts the onset of bimodality for $v_0\tau/S>1/8$.
From Fig.~\ref{fig:density_smallintermediateS}, we also observe that the increase of $v_0\tau/S$ beyond this threshold enhances the bimodality showing two symmetric peaks with increasing height but occurring at spatial positions which get closer.

In the large-persistence regime $v_0\tau/S\gg1$ (see Fig.~\ref{fig:densityCosinus_smallS}), we observe the emergence of many symmetric peaks in $\rho(x)$. % (see panel (c) and panel (d)). 
Their positions are still determined by the balance between $u(x)$ and $F(x)/\gamma$, and, in this case, roughly coincide with the minima of $u(x)$ close to the origin (i.e., the minimum of $U(x)$). % (as shown Fig.~\ref{fig:densityCosinus_smallS}~(a) and~(b)).
As shown in Fig.~\ref{fig:densityCosinus_smallS}~(a), $u(x)$ first crosses $F(x)/\gamma$ almost in its first minima (at $x/v_0\tau \approx \pm0.15$) for $v_0\tau/S=4$.
This implies that small fluctuations of the self-propulsion velocity allows the particle to explore spatial regions which are even more distant from the potential minimum, 
so that it also accumulates at the second crossing point % between $u(x)$ and $F(x)/\gamma$ 
(at $x/v_0\tau \approx 0.4$). 
According to Fig.~\ref{fig:densityCosinus_smallS}~(c), the height of these secondary peaks is smaller than that of the primary ones because the particle remains trapped at the first balance points for most of the time,
while only on rare occasions its swim velocity is sufficient to further climb up the potential gradient.
%majority of the particles remains trapped at the first balance point, while only a smaller fraction of them has (on average) a swim velocity sufficient to explore further spatial regions.
%In Fig.~\ref{fig:densityCosinus_smallS}~(d), for an even larger value of $v_0\tau/S$, we observe the opposite situation. In this case, since the first minimum of $u(x)$ is still larger than $F(x)/\gamma$, the two symmetric peaks nearest to the origin are lower than the successive peak, where the first intersection of $u(x)$ and $F(x)/\gamma$ occurs.
In Fig.~\ref{fig:densityCosinus_smallS}~(d), for an even larger value of $v_0\tau/S=10$, we observe that the height of the peaks near the origin is lower than that of the successive peaks. 
In this case, Fig.~\ref{fig:densityCosinus_smallS}~(b) shows that the minima of $u(x)$ closest to the origin are still larger than $F(x)/\gamma$, 
so that (most of the time) % (on average) 
the particle has a sufficiently large self-propulsion velocity to go further until entering the spatial region where the first intersection of $u(x)$ and $F(x)/\gamma$ occurs.
We conclude that, even in the case of a harmonic potential, the oscillation of the swim velocity allows the AOUP to climb up the potential barrier and accumulate preferredly in spatial regions (corresponding to minima of $u(x)$),
%, also accumulating in the minima of $u(x)$ 
which are further away from the origin. 
%{\rene We stress that, while it is generally expected that active particles accumulate near the minima of $u(x)$ in the potential-free case, we predict here,  an additional bias towards accumulation in the first minimum (relative to the minimum of the potential)
%at which the particle sufficiently experiences the confining force $F(x)/\gamma\geq u(x)$.
%In other words, a strong spatial oscillation of the swim velocity allows the AOUP to climb up the potential barrier, such that $\rho(x)$ exhibits two distinct high maxima far away from the origin.}
% This observation is, on the large scale, distinct from the behavior of an AOUP with constant swim velocity, which is most likely be found in the center of a harmonic trap. Due to the non-Gaussian behavior induced by $u(x)$, the generalized AOUP model considered here might be even more similar to the ABP model with a space-dependent self propulsion than in the special case with $u(x)=v_0$.
%they climb on the potential barrier (concentrating around minima far from the origin) and, thus, prefer staying far away from the minimum of the potential.
%{\rene\it[I find this point pretty important, maybe we should move it to or repeat it in the conclusions!?!]}

Finally, we note that in the large-persistence regime, the UCNA prediction~\eqref{eq:UCNA_generalized} for the spatial distribution fails. 
This occurs because of the presence of spatial regions where the effective friction $\Lambda(x)$, given by Eq.~\eqref{eq:lambda_harmonic}, becomes negative (see the gray-shaded regions in Fig.~\ref{fig:densityCosinus_smallS}(c) and~(d)).
This implies that also the corresponding approximation for $\rho(x)$ can assume negative values.
 This failure resembles the one of the UCNA (or the Fox approach) for the standard AOUP model with $u(x)=v_0$ confined in a nonconvex potential \cite{wittmann2017effective}. 
In that case, the strongly non-Gaussian nature of the system is at the basis of new intriguing phenomena, such as the occurring of effective negative mobility regions~\cite{caprini2018activeescape}, the overcooling of the system~\cite{schwarzendahl2021anomalous} and the violation of the Kramers law for the escape properties~\cite{caprini2021correlated, woillez2019activated}.
We expect that our model could display a similar phenomenology and that such problems can be treated by using similar theoretical techniques~\cite{caprini2018activeescape, fily2019self, woillez2020active, woillez2020nonlocal}. 
However, we stress that the generalized UCNA still accurately predicts the positions of the main peaks in Fig.~\ref{fig:densityCosinus_smallS}~(c),
although  in Fig.~\ref{fig:densityCosinus_smallS}~(d) there emerge additional  smaller peaks further away from the origin, which are absent in simulations.
The appearance of those fake peaks is reminiscent of the overestimated %can be attributed to the same overestimate 
wall accumulation predicted by UCNA for $u(x)=v_0$.
%{\rene Those fake peaks might be of similar origin than the overestimated the effective wall accumulation predicted by UCNA for $u(x)=v_0$ [...].}
%{\color{red} I don't think so (or maybe I don't understand what you mean). Here, it seems that the particles can explore further spatial regions. For $u(x)=v_0$ the main peaks of the simulation roughly coincides with that of UCNA.}

%{\rene\it Just an idea: can we avaluate the theory in the limit $v_0\tau/S \gg 1$ and only look at the maxima of $\rho(x)$, maybe even in the high-activity limit??}

%---------------------------- Fig.2 ----------------------------------
\begin{figure*}[t!]
\includegraphics[width=1\textwidth]{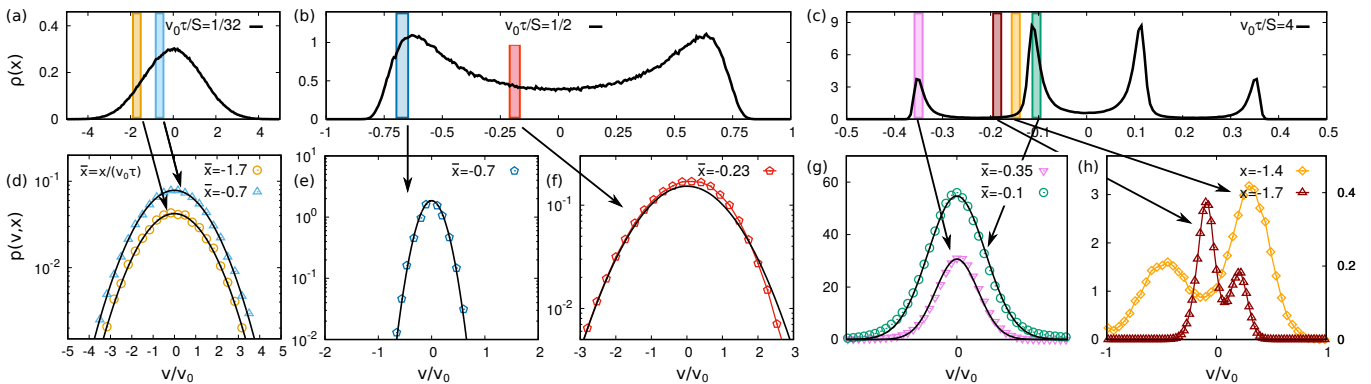}
\caption{
Velocity distributions. Panels (a), (b) and (c): density distribution $\rho(x)$ for $S=32, 2, 0.25$, respectively. Colored rectangles are drawn in correspondence of the spatial regions used to calculate the velocity distribution in the other panels.
Panels (d), (e), (f), (g) and (h): distribution $p(v,x)$ as a function of $v$ calculated at fixed positions $x$. Panel (d) is calculated at $S=32$, panels (e) and (f) at $S=2$ and panels (g) and (h) at $S=0.25$. 
Colored points are obtained by numerical simulations while solid black lines by plotting the theoretical prediction~\eqref{eq:powerexpansionintau}. 
Simulations are realized with $\tau k/\gamma=1$ and $\alpha=0.9$. 
 }
\label{fig:vel_distr_largeInterS}
\end{figure*}
%---------------------------------------------------------------------

\subsection{Velocity distribution}

We now focus on the dependence of the full joint probability density $p(x,v)$  on the velocity, shown in %study the properties of the velocity distribution at a fixed the particle position. 
Fig.~\ref{fig:vel_distr_largeInterS} %shows the distribution $p(x,v)$ calculated
for some representative values of the particle's position $x$. 
%In particular, 
Moreover, we choose %we evaluate 
three different values of $S/(v_0\tau)$ to explore the three distinct regimes observed in Sec~\ref{subsec:density}.
For each regime, we report once again the density distribution $\rho(x)$ (panels (a), (b) and (c)), where colored bars %rectangles
mark the regions for which we calculate $p(x,v)$ as a function of $v$ in panels (d), (e), (f), (g), (h).

In the regime of small persistence, $(v_0\tau)/S \ll 1$, the shape of $p(v,x)$ is Gaussian, independently of the position $x$ (Fig.~\ref{fig:vel_distr_largeInterS}~(d)). 
This result fully agrees with the prediction~\eqref{eq:powerexpansionintau} as revealed by the comparison between colored data points and black solid lines in Fig.~\ref{fig:vel_distr_largeInterS}~(d). 
As predicted by the space-dependent variance in Eq.~\eqref{eq:spatialvariance}, different positions $x$ come along with a change in the width of the velocity distribution.

%The regime of intermediate persistence, $v_0\tau/S \sim 1$, which displays a bimodality in the density distribution, also reveals a richer phenomenology of % scenario also in the stationary velocity profile.
In Fig.~\ref{fig:vel_distr_largeInterS}~(e) and~(f), the regime of intermediate persistence, $v_0\tau/S \sim 1$, which displays a bimodality in the density distribution, is investigated.
Here, we compare $p(x,v)$ calculated in the vicinity of a %approximatively around a 
peak of $\rho(x)$ to the velocity profile near the local minimum of $\rho(x)$ (at the origin).
In the former case, the distribution $p(x,v)$ displays an almost Gaussian shape in agreement with Eq.~\eqref{fig:vel_distr_largeInterS}, 
while in the latter case, it deviates from the theoretical prediction and shows a non-Gaussian nature. 
In particular, the shape of $p(x,v)$ becomes asymmetric in $v$ and develops non-Gaussian tails. 
%{\rene (not shown in Fig.~\ref{fig:vel_distr_largeInterS}~(f) [???])}.
While the prediction~\eqref{fig:vel_distr_largeInterS} cannot account for non-Gaussianity, we remark that its quality near the regions where the particle preferably accumulates resembles the one obtained in Ref.~\cite{caprini2019activity}, where an AOUP (with $u(x)=v_0$) has been studied in a single-well anharmonic confinement.
%Also in that case, the active particle climbs on the external potential so that the density displays a bimodal behavior and the velocity dis while the velocity variance shows a space-dependence through the second derivative of the potential, $\Lambda^{-1}(x) = \Gamma^{-1}(x)$and space-dependent 
%Also in Ref.\cite{}, the non Gaussian nature of $p(v,x)$ ad a function of $v$ (at fixed $x$) emerges near the minimum of the potential.

Finally, the large-persistence regime, $v_0\tau/S \gg 1$, where the density distribution has multiple peaks  also gives rise to a rich phenomenology of the stationary velocity profile, as shown in Fig.~\ref{fig:vel_distr_largeInterS}~(g) and~(h).
In the spatial regions for which $\Lambda(x)>0$, i.e., where the particles accumulate, the velocity distribution $p(x,v)$ (at fixed $x$) is described by the Gaussian distribution with space-dependent variance given by Eq.~\eqref{fig:vel_distr_largeInterS} (see Fig.~\ref{fig:vel_distr_largeInterS}~(g)), as in the case $v_0\tau/S \lesssim 1$.
Instead,  in the spatial regions where $\Lambda(x)<0$, i.e., between the primary and the secondary peaks (see also Fig.~\ref{fig:densityCosinus_smallS}~(c)), the distribution displays a non-Gaussian shape (see Fig.~\ref{fig:vel_distr_largeInterS}~(h)).
Compared to the case $v_0\tau/S \sim 1$, the non-Gaussianity is much more evident due to %  revealing also 
the occurrence of a bimodal behavior in the velocity distribution. 
In more detail, upon shifting the position $x$ in the $p(x,v)$ closer to the origin (Fig.~\ref{fig:vel_distr_largeInterS}~(g) and~(h)), we observe
that, starting from a nearly Gaussian shape centered at $v=0$ (pink curve), the main peak moves toward $v<0$ and a second small peak starts growing for $v>0$ (brown curve).
%As the position $x$ gets closer to the origin, the main peak of the conditional velocity distribution moves away from $v=0$ and a second small peak starts growing for large values of $v$. 
Shifting again $x$, the second peak becomes dominant (yellow curve) and moves closer toward $v=0$ until the distribution is again described by a Gaussian (green curve). % in the spatial region where $\Lambda(x)>0$.
This phenomenology resembles the one observed in the case of 
an AOUP with $u(x)=v_0$ in a double-well potential~\cite{caprini2018activeescape}. 
Also in the latter case, the velocity distribution at fixed position exhibits bimodality in the spatial regions where the effective friction coefficient $\Lambda(x)\simeq\Gamma(x)$ becomes negative,
although this effect is then induced by the negative curvature of the potential. 
%, which leads to negative values of $\Lambda(x)=\Gamma(x)$.

\subsection{Spatial profile of the kinetic temperature}

%---------------------------- Fig.4 ----------------------------------
\begin{figure*}[t!]
\includegraphics[width=0.75\textwidth]{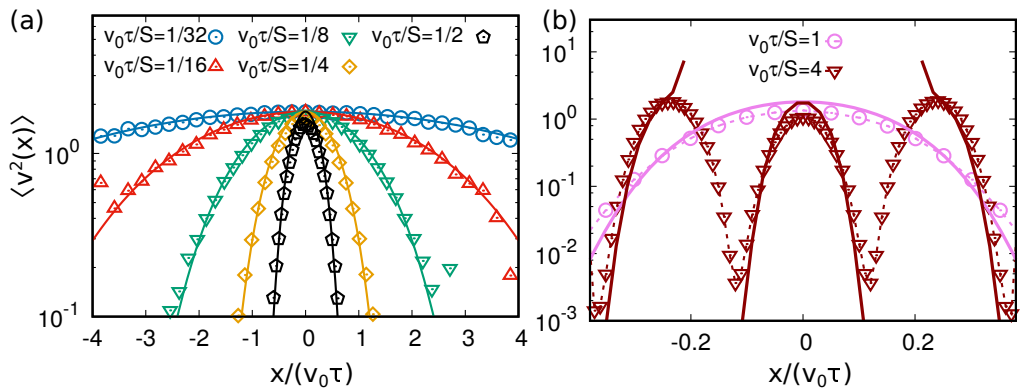}
\caption{
Spatial profile of the kinetic temperature. $\langle v^2(x) \rangle$ as a function of $x$ for different values of the dimensionless parameter $v_0\tau/S$ at fixed $\tau k /\gamma$ and $\alpha$.
Panel (a) shows the small-persistence regime, $v_0\tau/S \leq 1$, while panel (b) displays the intermediate-persistence and large-persistence regimes, namely $v_0\tau/S \sim 1$ and $v_0\tau/S \geq 1$, respectively.
Colored points are obtained from numerical simulations while solid colored lines from the theoretical prediction~\eqref{eq:spatialvariance}. Simulations are realized with $\tau k/\gamma=1$ and $\alpha=0.9$.
 }
\label{fig:temperatureprofile}
\end{figure*}
%---------------------------------------------------------------------

To emphasize the dynamical effects due to the spatial modulation of the swim velocity,
we focus on the profile of the kinetic temperature defined as the variance of the particle velocity,  $\langle v^2(x)\rangle$.
We show $\langle v^2(x)\rangle$ as a function of $x$ in Fig.~\ref{fig:temperatureprofile} for values of $v_0\tau/S$ spanning all regimes from small (panel~(a)) to intermediate and large persistence (panel~(b)).

For small values of $v_0\tau/S\ll1$, the profile $\langle v^2(x)\rangle$ is rather flat and attains its maximum value at $x=0$, i.e., the position of the potential minimum (see Fig.~\ref{fig:temperatureprofile}~(a)). 
When $v_0\tau/S$ increases, e.g., due to a shorter periodicity $S$ of the swim-velocity $u(x)$, the variance of the particle velocity becomes steeper and decreases to zero in the regions which are not explored by the particle. %decreases , the profile of $\langle v^2(x)\rangle$ and varies more steeply. 
This is consistent with the scenario observed in Fig.~\ref{fig:density_smallintermediateS}~(c): the particles accumulate in the regions where they move slowly and the velocity variance is small. 
Such a result agrees with the observed behavior in the potential-free case (such that $u(x)$ coincides the particle velocity),
where the particles accumulate in regions corresponding to the minima of $u(x)$, according to the law $\rho(x)\sim 1/u(x)$. 
In this regime, the comparison between numerical data and the theoretical prediction~\eqref{eq:spatialvariance} (with $\Lambda(x)$ given by Eq.~\eqref{eq:lambda_harmonic}) shows a good agreement. 

For larger values of $v_0\tau/S$ (large-persistence regime), the velocity variance shows a more complex profile (see Fig.~\ref{fig:temperatureprofile}~(b)), which resembles the oscillating shape of $u(x)$. 
In particular, $\langle v^2(x) \rangle$ is very small near the peaks of the density distribution, while assumes larger values in the regions where the density is very small and the probability of finding a particle is very low.
This finding is consistent with the fact that active particles accumulate in the regions where the velocity variance is small.
Finally, in this regime, the prediction~\eqref{eq:spatialvariance} reproduces quite well the behavior near the origin but fails further away from it, 
specifically, in the regions where the effective friction displays negative values, $\Lambda(x)<0$.

\section{Conclusion}\label{sec:conclusion}

%In this paper, we have investigated the stationary behavior of an active particle subject to two kind of spatially-dependent drivings: both the self-propulsion force and the external force vary with the particle's position. 
In this paper, we have investigated the stationary behavior of an active particle subject to two competing spatial-dependent  drivings: the self-propulsion velocity and the external force.
While the two mechanisms were already investigated separately, to the best of our knowledge, this is the first time
that their interplay has been considered. 
Starting from a Fokker-Planck description of the particle's dynamics in our generalized AOUP model~\cite{caprini2021spatial}, we have developed a theoretical treatment, 
applicable to rather general choices of confining potentials and inhomogeneous swim velocities which provides the steady-state distribution~\eqref{eq:powerexpansionintau} of both positions and velocities as a function of the input potential and of the swim velocity profile.
The theory presented here contains as special cases
both  the UCNA describing the time evolution of distribution of positions and velocities of an AOUP with constant swim velocity in an external field \cite{maggi2015multidimensional, marconi2017heat},
%ii) the UCNA describing the time evolution of distribution of positions of an AOUP with space-dependent swim velocity (in its original version for a particle subject to a multiplicative Gaussian colored noise) in an external field,
%BETTER OTHER WAY ROUND: ... which has only been achieved in a different context of a multiplicative Gaussian colored noise
%iii) the stationary solution in the Fox approach (which is only equal to the UCNA result in the stationary case) for the same setup %  describing the stationary positional distribution
%and 
and the recent theory of a free AOUP driven by an inhomogeneous propulsion force, which is exact in the stationary case~\cite{caprini2021dynamics}.
%The method is based based on a perturbative analysis of the Fokker-Planck-Kramers employing the persistence time as small parameter.
%The predictions of the  perturbative expansion concerning  the density profile as well as the velocity distribution and the kinetic temperature profile become exact in the small persistence limit and provide useful approximations in the large persistence regime.
 % 
 Our theoretical method is exact in the small-persistence regime, being consistent with the results obtained through an exact perturbative method, and also provides a useful approximation to qualitatively predict the shape of the distributions in the large-persistence regime.

Specifically, we have applied our theory to an AOUP in a sinusoidal motility landscape subject to a harmonic potential. 
We corroborated all theoretical predictions by numerical simulations and found a good agreement. 
The system revealed an intriguing scenario determined by the joint action of the self-propulsion velocity gradient and the external force.
While, in the regime of small persistence, both the density and the velocity distributions are bell shaped and well-approximated by Gaussians, we predict that, as the persistence length becomes comparable with the spatial period of the swim velocity, a transition from a unimodal to a bimodal density occurs, also accompanied by a strong non-Gaussian effects in the velocity distribution. 
Interestingly, in the large-persistence regime, as the density shows multi-modality, the velocity distribution becomes bimodal in the spatial regions between two successive peaks of the density.

%Regarding future applications and developments it is possible by the same methods to investigate different types of setups, such as more general potentials and swim-velocity profiles and, by incorporating suitable extensions of the UCNA, to study the same physics in higher dimensions. 

Moreover, we have shown that the interplay between the external force and the modulation of the swim velocity can be used to manipulate the behavior of a confined active particle, for instance by locally increasing the kinetic temperature or by forcing the particles to accumulate in distinct spatial regions with different probability.
This possibility to fine-tune the steady-properties of active particles opens up a new avenue for future applications and developments.
While, for active colloids, the emergence of an additional effective torque due to the spatial modulation of the swim velocity could be responsible for an even more complex phenomenology~\cite{lozano2016phototaxis, jahanshahi2020realization}, we outline that our theory should be suitable in the case of engineered bacteria whose velocity profile can be manipulated by external light~\cite{arlt2018painting, frangipane2018dynamic}.
%As a next step our model can be used to study mechanical properties~\cite{wittmann2019pressure} and collective phenomena~\cite{...} in such systems. {\color{red}\it[could you add a few more citations here?]}
From a pure theoretical perspective, our techniques may also be extended and applied to more complex dynamics, for instance accounting for the presence of thermal noise, a space-dependent torque~\cite{lozano2016phototaxis, jahanshahi2020realization}, or additional competing nonconservative force fields like a Lorentz force \cite{vuijk2020lorentz}.
%For these problems, we anticipate that the stationary...}
%In the latter case, it is well-known that UCNA  developed so far and the Fox approach differ in the case of constant swim velocity~\cite{...} [CIT.]. 
%A detailed comparison between them also in the case of spatial dependent swim velocity could shed light on the relation between the two approaches. 
%
A final challenging research point concerns the dynamical properties of our model and, in particular, the extension of the theory to time-dependent swim velocity profiles $u(x,t)$, for instance in the form of traveling waves~\cite{lozano2019propagating, koumakis2019dynamic, geiseler2017self}.

\section*{Acknowledgements}
LC and UMBM warmly thank Andrea Puglisi for letting us use the computer facilities of his group and for discussions
regarding some aspects of this research. 

% TODO: include author contributions
%\paragraph{Author contributions}
%This is optional. If desired, contributions should be succinctly described in a single short paragraph, using author initials.

% TODO: include funding information
\paragraph{Funding information}
LC and UMBM acknowledge support from the MIUR PRIN 2017 project
201798CZLJ. LC acknowledges support from the Alexander Von Humboldt foundation. RW and HL acknowledge
support by the Deutsche Forschungsgemeinschaft (DFG) through the SPP 2265, under grant numbers WI 5527/1-1
(RW) and LO 418/25-1 (HL).
%Authors are required to provide funding information, including relevant agencies and grant numbers with linked author's initials. Correctly-provided data will be linked to funders listed in the \href{https://www.crossref.org/services/funder-registry/}{\sf Fundref registry}.

\appendix

\section{Derivation of the auxiliary dynamics~\eqref{eq:dynamics_vel}}
\label{appendix_1}
To derive the auxiliary dynamics~\eqref{eq:dynamics_vel}, we start from Eq.~\eqref{eq:AOUP_x} chosing $D_\text{t}=0$.
We recall that following Ref.~\cite{caprini2018active} it is possible to generalize the procedure also to include the more general case with $D_\text{t}>0$.
At first, we take the time-derivative of Eq.~\eqref{eq:AOUP_x}, obtaining:
\begin{equation}
\gamma \ddot{\mathbf{x}}= -\nabla \nabla U \cdot \dot{\mathbf{x}} +  \gamma \boldsymbol{\eta} \left( \frac{\partial}{\partial t} + \mathbf{v}\cdot\nabla\right) u(\mathbf{x}, t) + \gamma u(\mathbf{x}, t)\dot{\boldsymbol{\eta}} \,.
\end{equation}
By defining the $\mathbf{v}=\dot{\mathbf{x}}$ as the particle velocity and replacing $\dot{\mathbf{f}}_a$ with the dynamics \eqref{eq:AOUP_eta}, we get:
\begin{equation}
\begin{aligned}
\gamma\dot{\mathbf{v}}=&-\nabla \nabla U \cdot {\mathbf{v}}  +  \gamma \boldsymbol{\eta} \left( \frac{\partial}{\partial t} + \mathbf{v}\cdot\nabla\right) u(\mathbf{x}, t)+ \gamma u(\mathbf{x}, t)\left(-\frac{\boldsymbol{\eta}}{\tau} + \frac{\sqrt{2}}{\sqrt{\tau}}\boldsymbol{\chi} \right) \,.
\end{aligned}
\end{equation}
Finally, by replacing $\boldsymbol{\eta}$ in favor of $\mathbf{v}$ and $\mathbf{x}$, taking advantage of the relation~\eqref{eq:AOUP_x}, we obtain the dynamics~\eqref{eq:dynamics_vel}.

%%%%%%%%%%%%%%%%%%%%%%%%%%%%%%%%%%%%%%%%%%%%%%%%%%%%%
\section{Derivation of predictions~\eqref{eq:powerexpansionintau} and~\eqref{eq:UCNA_generalized}}
\label{appendix_2}

To predict the shape of the stationary probability distributions, $p(\mathbf{x}, \mathbf{v})$ and $\rho(\mathbf{x})$, stated in Sec.~\ref{sec:Theoretical predictions},
we start from the dynamics in the variables $\mathbf{x}$ and $\mathbf{v}$, namely Eq.~\eqref{eq:dynamics_vel},
for a static profile of the swim velocity, $u(\mathbf{x})$.
Switching to the Fokker-Planck equation for $p=p(\mathbf{x}, \mathbf{v}, t)$, we obtain:
\begin{equation}
\begin{aligned}
\partial_t p=&\nabla_v \cdot \left(\frac{\boldsymbol{\Gamma}}{\tau}\cdot \mathbf{v} p + \frac{u^2(\mathbf{x})}{\tau}\nabla_v p  \right) -\mathbf{v}\cdot\nabla p + \nabla \cdot \frac{\nabla U}{\gamma\tau} p- \nabla_v \cdot\frac{\left[ \gamma\mathbf{v} + \nabla U \right]}{\gamma u(\mathbf{x})} \left( \mathbf{v} \cdot \nabla \right) u(\mathbf{x}) \,,
\end{aligned}
\end{equation}
where $\nabla$ %is the spatial divergence 
and $\nabla_v$ %=(\partial_{v_x}, \partial_{v_y}, ...)$.  
are the vectorial derivative operators in position and velocity space, respectively.
Balancing the diffusion term (proportional to the Laplacian of $\mathbf{v}$) and the other effective friction terms (say the one linearly proportional to $\mathbf{v}$), we get the condition:
\begin{equation}
\label{eq:irreversible_currents}
0= \nabla_v \cdot \left(\frac{\boldsymbol{\Lambda}}{\tau}\cdot \mathbf{v} p + \frac{u^2(\mathbf{x})}{\tau}\nabla_v p  \right)  \,
\end{equation}
%{\rene for a (current-free) stationary state with
with the effective friction matrix
\begin{equation}
\label{eq:general_lambda_def_app}
\boldsymbol{\Lambda}(\mathbf{x}) = \boldsymbol{\text{I}} +\frac{\tau}{\gamma}\nabla \nabla U (\mathbf{x}) - \frac{\tau}{\gamma} \nabla U(\mathbf{x})\frac{\nabla u(\mathbf{x})}{u(\mathbf{x})} \,,
\end{equation}
that has been defined in Eq.~\eqref{eq:general_lambda_def}.
The condition~\eqref{eq:irreversible_currents} corresponds of requiring that all the irreversible currents are zero, in the same spirit of Ref.~\cite{marconi2017heat}. The solution of Eq.~\eqref{eq:irreversible_currents} corresponds to Eq.~\eqref{eq:powerexpansionintau}, where $\rho(\mathbf{x})$ is still an unknown function to be determined below. 

To determine the function $\rho(\mathbf{x})$, we first identify the
 acceleration term~\cite{jung1988multi}
\begin{equation}
u(\mathbf{x})\frac{d}{dt} \frac{\mathbf{v}}{u(\mathbf{x})}=
\dot{\mathbf{v}}-\frac{ \mathbf{v}  }{ u(\mathbf{x}, t)} \left(  \mathbf{v} \cdot \nabla \right) u(\mathbf{x}, t)  \,
\end{equation}
in the dynamics \eqref{eq:dynamics_vel}.
Assuming that the velocity relaxes faster than the position allows us to neglect both these terms in Eq.~\eqref{eq:dynamics_vel}, obtaining the following overdamped equation:
\begin{equation}
%\dot x_i=-\frac{1}{\gamma}\Lambda^{-1}_{ij} \nabla_j U +v_0(x) \Lambda^{-1}_{ij} \sqrt{2\tau}\chi_j
\dot{\mathbf{x}}=-\frac{1}{\gamma}\boldsymbol{\Lambda}^{-1}\cdot \nabla U +\sqrt{2\tau}u(\mathbf{x}) \boldsymbol{\Lambda}^{-1}\cdot\boldsymbol{\chi} \,.
\end{equation}
From this dynamics, it is convenient to switch to the effective Smoluchowski %Fokker-Planck
equation for the density of the system, $\rho(\mathbf{x},t)$, and use the Stratonovich convention,  obtaining: 
\begin{eqnarray}&&
\frac{\partial \rho}{\partial t}=\frac{\partial}{\partial x_i}  \left(\frac{1}{\gamma}\Lambda^{-1}_{ij} \left(\frac{\partial U}{\partial x_j} \right)  \, \rho +\tau 
u \Lambda^{-1}_{ik}  \frac{\partial}{\partial x_j} \left[\Lambda^{-1}_{jk} u  \rho\right]\right) \,.
\label{eq:effectiveUCNA}
\end{eqnarray}
%where we restricted to the case $u=u(\mathbf{x})$ which admits a time-independent steady-state.
Here and in what follows, we have explicitly written Latin indices for the spatial components of vectors and matrices and adopted also the Einstein's convention for repeated indices, for convenience.

To proceed, we assume the zero-current condition, obtaining an effective equation for the stationary density $\rho(\mathbf{x})$:
\begin{equation}
 \frac{1}{\gamma}\Lambda^{-1}_{ij} \left(\frac{\partial U}{\partial x_j}\right)\, \rho +\tau 
u \Lambda^{-1}_{ik}  \frac{\partial}{\partial x_j} \left[\Lambda^{-1}_{jk} u  \rho\right]=0 \,.
\end{equation}
%
%Multiplying by $\Lambda_{li} $ and summing over $i$, we get the following relation after some algebraic manipulations
%\begin{equation}
%\frac{1}{\gamma}\Lambda_{li} \Lambda^{-1}_{ij} \nabla_j U  \, \rho +\tau 
%u \Lambda_{li}\Lambda^{-1}_{ik}  \frac{\partial}{\partial x_j} [\Lambda^{-1}_{jk} u  \rho]=0
%\end{equation}
%then
%\begin{equation}
% \frac{1}{\gamma}\nabla_l U  \, \rho +\tau u \delta_{lk}  \frac{\partial}{\partial x_j} [\Lambda^{-1}_{jk} u  \rho]=0
%\end{equation}
% \begin{equation}
%\frac{1}{\gamma} \nabla_l U  \, \rho +\tau u   \frac{\partial}{\partial x_j} [\Lambda^{-1}_{jl} u  \rho]=0
%\end{equation}
%
%\begin{equation}
%\frac{1}{\gamma \tau u^2} \nabla_l U  \, u \rho +    \frac{\partial}{\partial x_j} [\Lambda^{-1}_{jl} ] u  \rho
% + \Lambda^{-1}_{jl}   \frac{\partial}{\partial x_j} [u  \rho]=0
%\end{equation}
%%%
%{\color{red} The following steps assume the symmetry of $\Lambda$ which does not always hold}.\\
%Multiply by $\Lambda_{il}  $ 
%
%{\rene\it[I changed the following equations significantly due to several typos and problems, please check again carefully!!!]}
Multiplying by $\Lambda_{lh}\Lambda_{hi}$ and summing over repeated indices, we get the following relation after some algebraic manipulations
\begin{equation}
\frac{1}{\gamma \tau u^2} \Lambda_{lj}  \frac{\partial U}{\partial x_j}  +  \Lambda_{lk}   \frac{\partial}{\partial x_j} \Lambda^{-1}_{jk} 
 +\frac{\Lambda_{lk}  \Lambda^{-1}_{jk}}{u  \rho}   \frac{\partial}{\partial x_j} [u  \rho]=0 \,,
\label{eq:rhostat}
\end{equation}
%we get the following equation:
%
%\begin{equation}
%\frac{1}{ u \rho} \frac{\partial}{\partial x_l} [u  \rho]= -\frac{1}{\gamma \tau u^2} \Lambda_{lj}  \frac{\partial U}{\partial x_j}  \, +   \frac{\partial}{\partial x_l} \ln( \det \Lambda) \,,
%\frac{1}{ u \rho} \frac{\partial}{\partial x_l} [u  \rho]= -\frac{1}{\gamma \tau u^2} \Lambda_{lj}  \frac{\partial U}{\partial x_j}  -  \Lambda_{lk}   \frac{\partial}{\partial x_j} \Lambda^{-1}_{jk} \,,
%\label{eq:rhostat}
%\end{equation}
%whose solution for the density distribution $\rho(\mathbf{x})$ is given by Eq.~\eqref{eq:UCNA_generalized}.
whose solution for the density distribution $\rho(\mathbf{x})$ reads:
\begin{equation}
\label{eq:app_solution_UCNA_general}
\rho(\mathbf{x}) \approx \frac{\mathcal{N}}{u(\mathbf{x})}  \exp{\Biggl( \frac{1}{\gamma \tau} \int^x d\mathbf{y}\cdot \frac{\boldsymbol{\Lambda}(\mathbf{y}) \cdot\mathbf{F}(\mathbf{y}) }{u^2(\mathbf{y})}  + \int^x d\mathbf{y}\cdot \boldsymbol{\Lambda}(\mathbf{y}) \cdot \nabla \cdot \boldsymbol{\Lambda}^{-1}(\mathbf{y})\Biggr)} \,.
\end{equation}
Finally, by assuming a planar symmetry for both $u$ and $U$, we have $\nabla\cdot\boldsymbol{\Lambda}^{-1}\equiv \hat{e}_x \cdot \partial\boldsymbol{\Lambda}^{-1}/\partial x$,
where $\hat{e}_x$ denotes the unit vector corresponding to the coordinate $x$,
and can therefore use the explicit Jacobi relation
\begin{equation}
\boldsymbol{\Lambda}\cdot\hat{e}_x \cdot \frac{\partial\boldsymbol{\Lambda}^{-1}}{\partial x}
=-\frac{1}{\det[\boldsymbol{\Lambda}]} \hat{e}_x \frac{\partial\det[\boldsymbol{\Lambda}]}{\partial x}
=-\hat{e}_x\frac{\partial\ln\det[\boldsymbol{\Lambda}]}{\partial x}
\label{eq_detln}
\end{equation}
for the determinant $\det\boldsymbol{\Lambda}$ of a matrix $\boldsymbol{\Lambda}$.
We remark that the general relation
\begin{equation}
\boldsymbol{\Lambda}\cdot\nabla\cdot \boldsymbol{\Lambda}^{-1}
=-\nabla\ln\det[\boldsymbol{\Lambda}]
\label{eq_detlngen}
\end{equation}
only holds in the above planar case~\eqref{eq_detln} or for a constant swim velocity $u(\boldsymbol{x})=v_0$, see also appendix B of Ref.~\cite{wittmann2017effective}.
However, since there are no conceptual differences, we can plug the approximation~\eqref{eq_detlngen}
into the prediction~\eqref{eq:app_solution_UCNA_general} to obtain the compact representation~\eqref{eq:UCNA_generalized} of $\rho(\boldsymbol{x})$ in the main text.

%Finally, by assuming the symmetry $\boldsymbol{\Lambda}$, we can use the general relation
%\begin{equation}
%\boldsymbol{\Lambda} \cdot \nabla \cdot \boldsymbol{\Lambda}^{-1}=-\frac{1}{\det \boldsymbol{\Lambda}} \nabla \det \boldsymbol{\Lambda}
%\Lambda_{lk}   \frac{\partial}{\partial x_j} \Lambda^{-1}_{jk} =-\frac{1}{\det \Lambda} \frac{\partial }{\partial x_l} \det \Lambda \,,
%\end{equation}
%for the determinant $\det \boldsymbol{\Lambda}$ of a matrix $\boldsymbol{\Lambda}$. 
%Plugging this identify into the solution~\eqref{eq:app_solution_UCNA_general} allows us to obtain the prediction~\eqref{eq:UCNA_generalized} of the main text.
%We remark that the matrix $\boldsymbol{\Lambda}$ is symmetric only under specific choices of $u(\mathbf{x})$ and $U(\mathbf{x})$. 
%Anyway, if $\boldsymbol{\Lambda}$ is not symmetric there are no conceptual differences and we can simply replace Eq.~\eqref{eq:UCNA_generalized} by Eq.~\eqref{eq:app_solution_UCNA_general}.

The same stationary condition~\eqref{eq:rhostat} can be obtained using the Fox approach~\cite{fox1986b} (when generalized to multiple components~\cite{rein2016,sharma2017}), 
while the corresponding time evolution differs from the UCNA dynamics~\eqref{eq:effectiveUCNA} by the additional occurrence of the factors $\Lambda^{-1}_{ij}$ and $\Lambda^{-1}_{ik}$ therein.
Note that, if we do not neglect the thermal Brownian noise in Eq.~\eqref{eq:AOUP}, also the stationary predictions of Fox and UCNA differ, even for a spatially constant swim velocity~\cite{wittmann2017effective}.

%\begin{equation}
%\frac{1}{\gamma \tau u^2} \Lambda_{il}  \nabla_l U  \, u \rho +  \Lambda_{il}   \frac{\partial}{\partial x_j} [\Lambda^{-1}_{jl} ] u  \rho
% +\Lambda_{il}  \Lambda^{-1}_{jl}   \frac{\partial}{\partial x_j} [u  \rho]=0 \,.
%\end{equation}
%By accounting for the symmetry of $\Lambda$, we obtain
%
%\begin{equation}
%\frac{1}{\gamma \tau u^2} \Lambda_{il}  \nabla_l U  \, u \rho +  \Lambda_{il}   \frac{\partial}{\partial x_j} [\Lambda^{-1}_{jl} ] u  \rho
% + \frac{\partial}{\partial x_i} [u  \rho]=0 \,,
% \label{equationrho}
%\end{equation}
%which, by using the definition of the determant of a matrix, reads:
%\begin{equation}
%\Lambda_{il}   \frac{\partial}{\partial x_j} [\Lambda^{-1}_{jl} ] =-\frac{1}{\det \Lambda} \frac{\partial }{\partial x_i} \det \Lambda \,.
%\end{equation}
%Finally, we got the Smoluchowski equation for the density distribution
%
%\begin{equation}
%\frac{1}{ u \rho} \frac{\partial}{\partial x_i} [u  \rho]= -\frac{1}{\gamma \tau u^2} \Lambda_{il}  \nabla_l U  \, +   \frac{\partial}{\partial x_i} \ln( \det \Lambda) 
%\end{equation}
%whose solution is given by Eq.~\eqref{eq:UCNA_generalized}.

%%%%%%%%%%%%%%%%%%%%%%%%%%%%%%%%%%%%%%%%%%%%%%%%%%%%%%%%%%%%%%%%%%%%%%%%%
\section{Multi-scale technique: derivation of Eq.~\eqref{eq:pxv_pertubativetau}}
\label{appendix}

%%%%%%%%%%%%%%%%%%%%%%%%%%%%%%%%%%%%%%%%%%%%%%%%%%%%%%%%%%%%%%%%%%%%%%%%%
 In this appendix, we derive the perturbative result~\eqref{eq:pxv_pertubativetau} for the probability distribution $p(x,v)$ in the one-dimensional active system described by the Fokker-Planck equation~\eqref{eq:FokkerPlanck_potentialconfined}.
We adopt the multiple-time-scale technique, which is designed to deal with problems with fast and slow degrees of freedom.
In the regime of  small persistence time (where $\tau$ is the smallest time scale of the system), the dynamics~\eqref{eq:FokkerPlanck_potentialconfined} exhibits the separation of time scales: in this case, the particle velocity rapidly arranges according to its stationary distribution and the spatial distribution evolves on a slower time scale. 

To derive the multiple-time expansion, let us introduce the following dimensionless variables:
%{\rene\it[ $F$ and $\Gamma$ are already used as dimensional quantities!]}
\begin{eqnarray}&&
\tilde t= t\frac{v_0}{S}\,,
\\&&
X=\frac{x}{S} \,,
\\&&
V=\frac{v}{v_0} \,,
\\&&
%\tilde \tau=\frac{v_0}{S} \tau \,,\\&&
\tilde{F}(X)= - \frac{S} {v_0^2 \tau \gamma} \frac{\partial U(x)}{\partial x} \,,
\\&&
\tilde{\Gamma}(X) =1-  \tau^2\frac{v_0^2}{S^2} %\tilde \tau^2
\frac{\partial \tilde{F}(X)}{\partial X} \,,
\\&&
w(X)=\frac{u(x)}{v_0} \,,
\end{eqnarray}
and the small expansion parameter $\zeta^{-1}\propto \tau$, where
\begin{equation}
\zeta=\frac{S}{\tau v_0} \,
\end{equation}
 is the ratio between the spatial period of the swim velocity $S$ and the persistence length of the self-propulsion velocity $v_0\tau$. 
With our choice, a large (small) value of $\zeta$ corresponds to the small-persistence (large-persistence) regime. 
Now, we express the Fokker-Planck equation~\eqref{eq:FokkerPlanck_potentialconfined} in these variables and find:
\begin{eqnarray}&& 
\frac{\partial  P(X,V,\bar t)}{\partial \bar t} +V \frac{\partial }{\partial X}  P 
+\tilde{F}(X) \frac{\partial }{\partial V}  P + \frac{1}{\zeta} R(X) \frac{\partial }{\partial V} V  P
\nonumber\\&&
+\frac{1}{w(X)}\frac{\partial}{\partial X} w(X)   \frac{\partial}{\partial V} (V^2 P) 
=\zeta L_\text{FP} P\,,
\label{eq:kramers0bb}
\end{eqnarray}
where we have further introduced the operator
\begin{equation}
L_\text{FP}\equiv\frac{\partial}{\partial V} \Bigl(  V+ w^2(X)\frac{\partial}{\partial V}      \Bigr) 
\end{equation}
and the function
\begin{equation}
R(X)\equiv \Bigl[ \frac{\partial \tilde{F}}{\partial X}-\tilde{F}(X) \frac{1}{w(X)}\frac{\partial}{\partial X} w(X)   \Bigr] \,,
\end{equation}
for convenience.

To develop our perturbative solution, we notice that the local operator $L_\text{FP}$ is proportional to the inverse expansion parameter $\zeta$ in Eq.~\eqref{eq:kramers0bb}. 
We find that $L_\text{FP}$
has  the following integer eigenvalues:
\begin{equation}
\nu=0, -1, -2, \ldots
%0,-\nu,-2\nu ,..
\end{equation}
and the Hermite polynomials as eigenfunctions:
\begin{equation}
H_\nu(X,V)=\frac{(-1)^\nu}{\sqrt{2\pi}} \left(w(X)\right)^{(\nu-1)}  \frac{\partial^\nu}{\partial V^\nu}\, e^{-\frac{V^2 }{2 w^2(X)}} \,.
\end{equation}
Using these basis functions, we obtain the ansatz to write the solution of the partial differential equation as a linear combination:
\begin{equation} 
P(X,V,\tilde t)=
\sum_{\nu=0}^\infty  \phi_\nu(X,\tilde t) H_\nu(X,V) \,.
\label{eq:hermiteexpansion}
\end{equation}
Upon substituting the expansion~\eqref{eq:hermiteexpansion} in Eq.~\eqref{eq:kramers0bb} and replacing $L_\text{FP}$ by its eigenvalues,
we obtain the equation:
\begin{eqnarray}&&
-\zeta %L_\text{FP}
\sum_\nu \nu\, \phi_\nu H_\nu=\sum_\nu \frac{\partial \phi_\nu}{\partial \tilde t}H_\nu \nonumber\\&&
+\sum_\nu V H_\nu \frac{\partial}{\partial X}\phi_\nu+\sum_\nu  \phi_\nu V\frac{\partial}{\partial X} H_\nu
+\tilde{F}\sum_\nu \phi_\nu \frac{\partial}{\partial V} H_\nu  \nonumber\\&&
+\frac{w'}{w} \sum_\nu \phi_\nu\frac{\partial}{\partial V} V^2 H_\nu  + \frac{1}{\zeta}R\sum_\nu\phi_\nu\frac{\partial}{\partial V} V H_\nu \,,
\label{eq:lfp}
\end{eqnarray}
from which we must determine the unknown functions $\phi_\nu(x,t)$.

Now, instead of truncating arbitrarily the infinite series in Eq.~\eqref{eq:lfp} at some order $\nu$, we consider the multiple-time expansion which orders the series in powers of the small parameter $1/\zeta$.
In such a way we perform an expansion near the equilibrium solution.
To this end, each amplitude $\phi_\nu$ (apart from $\phi_0(X,\tilde t)$ which is  of order $\zeta^0$) is expanded in powers of $1/\zeta$ as:
\begin{equation}
\phi_\nu(X,\tilde t)=
\sum_{s=0}^{\infty} \frac{1}{\zeta^s} 
\psi_{s \nu}(X,\tilde t) \,.
\label{eq:doubleexpansion}
\end{equation}
Then, we replace the actual distribution $P(X,V, \bar t)$ by an auxiliary distribution $P_a= P_a(X,V,\bar t_0,\bar t_1,\bar t_2,..)$, which reads:
\begin{equation} 
 P_a=
\sum_{s=0}^{\infty} \frac{1}{\zeta^s} 
\sum_{\nu=0}^{\infty}  
\psi_{s \nu}(X,\bar t_0,\bar t_1,\bar t_2,..)  H_{\nu}(X,V) \,.
\label{eq:pn}
\end{equation}
This distribution depends on many time variables $\{\bar t_s\}$, associated with the perturbation order $s$, which are defined as $\bar t_s=\bar t/\zeta^s$.
The time derivative
with respect to $\bar t$ is then expressed as the sum of partial time-like derivatives:
\begin{equation}
\frac{\partial}{\partial \bar t}=\frac{\partial}{\partial \bar t_0}
+\frac{1}{\zeta} \frac{\partial}{\partial \bar t_1}
+\frac{1}{\zeta^2} \frac{\partial}{\partial \bar t_2}+ \ldots \,.
\label{eq:mult}
\end{equation}

Substituting the expansions~\eqref{eq:doubleexpansion} and~\eqref{eq:mult} into Eq.~\eqref{eq:lfp} one obtains
at each order $1/\zeta^s$ and for each Hermite function an equation involving the amplitudes $\psi_{s \nu}(X,\tilde t)$.
The perturbative structure of the resulting set of equations is such that the amplitudes $\psi_{s \nu}(X,\tilde t)$ can be obtained by the amplitudes of the lower order $(s-1)$.
In particular, we find the following equation for $\psi_{00}=\phi_0$
\begin{eqnarray}&&
\frac{\partial \psi_{00}(X,\bar t)}{\partial \tilde t}=\frac{1}{\zeta}\frac{\partial}{\partial X} \Bigl[   
w(X) \frac{\partial }{\partial X}   \Bigl(w(X) \psi_{00}\Bigr)  \nonumber\\&&  -    
 \tilde{F}(X)   \psi_{00} +\frac{1}{\zeta^2} w(X) \frac{\partial }{\partial X} \Bigl(  w(X) R  \psi_{00}   \Bigr)  \Bigr] \,,
\label{equazione99a}
\end{eqnarray}
whose steady-state solution reads
\begin{equation}
\label{eq:psi00}
\begin{aligned}
 \psi_{00}(X)= &\frac{\mathcal{N}}{w(X) }\,  \Bigl( 1-\frac{1}{\zeta^2}  R(X) \Bigr)\times\exp\left[\int^X dy \frac{\tilde{F}(y)}{w^2(y)}  \Bigl( 1-\frac{1}{\zeta^2}  R(y) \Bigr) \right]\,,
\end{aligned}
\end{equation} 
where $\mathcal{N}$ is a normalization factor.
In our perturbative procedure, all the remaining amplitudes are expressed in terms of the pivot function $\psi_{00}(X)$.
The steady-state amplitudes of the higher-order Hermite polynomials are given by:
\begin{eqnarray}&&
\psi_{22} =\frac{1}{2} R(X)  \psi_{00}  \\&&
%\psi_{33}=-\frac{1}{3}[u \frac{\partial}{\partial X} \psi_{22} +u'\psi_{22}-\frac{F(X)}{u} \psi_{22}]\\&&
\psi_{33}(X)=-\frac{1}{6}w(X) \psi_{00}(X) \frac{\partial}{\partial X} R(X)\\&&
 \psi_{42}= -  \frac{3}{2}   \frac{\partial }{\partial X} [w(X)  \psi_{33}]   +R(X) \psi_{22}\\&&
 \psi_{44}(X)=-\frac{1}{4}\Bigl(\frac{\partial }{\partial X} [w(X)  \psi_{33}] -R(X) \psi_{22}\Bigr) \,,
\end{eqnarray}
where we have reported only the nonvanishing coefficients for $s\leq4$.
Note that, if $\nu > s$, the coefficients $\psi_{s\nu}$ are always zero.

Once Eq.~\eqref{eq:psi00} and the coefficients of the double series~\eqref{eq:pn} have been determined, one returns to the 
original dimensional variables and obtains the perturbative result for $p(x,v)$ reported in Eq.~\eqref{eq:pxv_pertubativetau}.

%%%%%%%%%%%%%%%%%%%%%%%%%%%%%%%%%%%%%%%%%%

%%%%%%%%%%%%%%%%%%%%%%%%%%%%%%%%%%%%%%%%%%

%\reftitle{References}


%\externalbibliography{yes}
%\bibliography{your_external_BibTeX_file}
%\externalbibliography{yes}

\bibliography{SD}

%=====================================
% References, variant B: internal bibliography
%=====================================
%\begin{thebibliography}{999}
% Reference 1
%\bibitem[Author1(year)]{ref-journal}
%Author1, T. The title of the cited article. {\em Journal Abbreviation} {\bf 2008}, {\em 10}, 142--149.
% Reference 2
%\bibitem[Author2(year)]{ref-book}
%Author2, L. The title of the cited contribution. In {\em The Book Title}; Editor1, F., Editor2, A., Eds.; Publishing House: City, Country, 2007; pp. 32--58.
%\end{thebibliography}

% The following MDPI journals use author-date citation: Arts, Econometrics, Economies, Genealogy, Humanities, IJFS, JRFM, Laws, Religions, Risks, Social Sciences. For those journals, please follow the formatting guidelines on http://www.mdpi.com/authors/references
% To cite two works by the same author: \citeauthor{ref-journal-1a} (\citeyear{ref-journal-1a}, \citeyear{ref-journal-1b}). This produces: Whittaker (1967, 1975)
% To cite two works by the same author with specific pages: \citeauthor{ref-journal-3a} (\citeyear{ref-journal-3a}, p. 328; \citeyear{ref-journal-3b}, p.475). This produces: Wong (1999, p. 328; 2000, p. 475)

%%%%%%%%%%%%%%%%%%%%%%%%%%%%%%%%%%%%%%%%%%
%% Optional
%\sampleavailability{Samples of the compounds ...... are available from the authors.}

%% for journal Sci
%\reviewreports{\\
%Reviewer 1 comments and authors’ response\\
%Reviewer 2 comments and authors’ response\\
%Reviewer 3 comments and authors’ response
%}

%%%%%%%%%%%%%%%%%%%%%%%%%%%%%%%%%%%%%%%%%%
\end{document}